\documentclass[12pt,a4paper]{article}

\usepackage{amsthm}
\usepackage{amsmath}
\usepackage{natbib}
\usepackage[colorlinks,citecolor=blue,urlcolor=blue,filecolor=blue,backref=page]{hyperref}
\usepackage{graphicx,psfrag,epsf}
\usepackage{subfigure}
\usepackage{xcolor}
\usepackage{algorithm}
\usepackage{adjustbox}
\usepackage[noend]{algpseudocode}
\usepackage{natbib}
\usepackage{bbm, dsfont}
\usepackage{amsfonts}
\usepackage{siunitx}
\newcommand{\be}{\begin{equation}}
\newcommand{\ee}{\end{equation}}

\usepackage[paperheight=11in,paperwidth=9.5in]{geometry}

\begin{document}

\begin{center}
	\Large \bf  Two-stage Circular-circular Regression with Zero-inflation: Application to Medical Sciences
\end{center}
\begin{center}
	\textbf{Jayant Jha$^{\ast}$ Prajamitra Bhuyan$^{\dag}$  }\footnote{Both authors contributed equally to this paper.}\\
	$^{\ast}$Institut de Neurosciences des Syst\`emes,  Aix-Marseille University, Marseille\\
	$^{\dag}$Department of Mathematics, Imperial College London, London

\end{center}

\begin{abstract}
This paper considers the modeling of zero-inflated circular measurements concerning real case studies from medical sciences. Circular-circular regression models have been discussed in the statistical literature and illustrated with various real-life applications. However, there are no models to deal with zero-inflated response as well as a covariate simultaneously. The M\"obius transformation based two-stage circular-circular regression model is proposed, and the Bayesian estimation of the model parameters is suggested using the MCMC algorithm. Simulation results show the superiority of the performance of the proposed method over the existing competitors. The method is applied to analyse real datasets on astigmatism due to cataract surgery and abnormal gait related to orthopaedic impairment. The methodology proposed can assist in efficient decision making during treatment or post-operative care.
\end{abstract}


\section{Introduction}{\label{Intro}}
In many real-world experiments, measurements are taken in angles, and such variables are commonly labeled as `circular variables' or `directional variables' in the statistics literature \citep{mardiabook00}. The need for developing statistical methods to study
circular regression is important due to its wide application in various fields of science, e.g., orthopaedics \citep[p-4]{Circ_Stat}, meteorology \citep{Kato08}, ecology \citep{fisher92665677}, and geoscience \citep{rivest97318324}. \citet{sarma93} proposed a generic circular-circular regression by joint modeling of the sine and cosine of the angular response on the trigonometric polynomial function of the angular covariate. In this context, some rotational models have been proposed where the predicted mean direction of the response is a fixed rotation of the covariate. See \citet{mackenzie576162} and \citet{rivest97318324} for more details. 
\citet{bhattacharya091433} proposed a Bayesian hierarchical framework for linear–circular regression.  %\citet{gould69683700} proposed regression models for the case of linear-circular regression. These models cannot be directly applied to the case of circular-circular regression due to the difference in the topology of a circle and Euclidean space. \citet{downs02683697}  proposed the M\"obius transformation based regression link function for circular-circular regression \textcolor{blue}{and considered it to study the relationships between the peak systolic blood pressure times on different days and wind directions at two different times. Later, \citet{Kato08} also considered the M\"obius transformation based link function by reparameterizing the model of \citet{downs02683697} and applied it to regress the wind directions on two different times and to study the association between the spawning time of fish with the time of low tide. Some of the other applications of the M\"obius transformation based circular-circular regression model are considered in opthalmology \citep{Jhatr1705} and gene-expression \citep{rueda150120}.} 
\citet{downs02683697} proposed the M\"obius transformation based regression link function for circular-circular regression to analyze the relationships between the peak systolic blood pressure times on different days, and wind directions at two different times. Later, \citet{Kato08} also considered the M\"obius transformation based link function by reparameterizing the model of \citet{downs02683697} and applied this model to study the association between the spawning time of fish and the time of low tide. Other applications of the M\"obius transformation based circular-circular regression model in opthalmology and genetics are discussed in \citet{Jhatr1705} and \citet{rueda150120}, respectively. In this article, we consider the problem of modeling circular measurements motivated by the following case studies from orthopaedics and ophthalmology.

\subsection{Abnormal Gait Data}\label{Ortho_data}
The ability to walk safely and efficiently is essential for an independent and productive life. Gait disorders lead to a loss of personal freedom, and increased risk of injuries that result in a marked reduction in the quality of life \citep{Gait_life}. In the last couple of decades, due to the increased awareness of the importance of gait control, researchers across the globe focused their research programs on the clinical and therapeutic aspects of walking and balance \citep{Gait_research}. Orthopaedic impairments of the lower extremities are the most common reasons for non-neurological gait disorders in adults \citep{Gait_life}. Patients with gait disorders caused by orthopaedic impairments are often referred to a physical therapist for treatment. Therapists visually observe the patient's gait to determine effective treatment protocol or evaluate the effect of a therapeutic intervention. Various kinematic features like sagittal plane angular movements (flexion-extension) at the ankle and knee are extremely important in determining abnormal gait Therapists visually observe the patient's gait to determine effective treatment protocol or evaluate the effect of a therapeutic intervention. Various kinematic features like sagittal plane angular movements (flexion-extension) at the ankle and knee are extremely important in determining abnormal gait\citep{Gait_data}.  Bio-engineers are interested to analyse the interlimb coordination concerning the flexion-extension to make an efficient design for the body-powered knee-ankle prosthesis \citep{Ankle_prosth}. The human walking cycle consists of a period when the foot is on the ground, followed by the next phase of the forward movement known as swing. Recent research suggests that the ankle extensors provide necessary kinetic energy for the initiation of swing \citep{Ankle_speed}. However, there is a lack of knowledge about the biomechanical process, which leads to the high power output of the ankle extensors \citep{Ankle_exten}. In this article, we consider a study conducted at the Bio-Engineering Unit, University of Calcutta, on a total of 29 individuals, including 10 healthy persons and 19 patients with unilateral orthopaedic impairment. Measurements on ankle and knee flexion-extension were recorded using electrogoniometers while subjects were walking at their own pace along an 8-meter long walkway turning around at each end. Previous research found no statistically significant difference between the motions of the right and left limb of a healthy person. Therefore, the angular movements of patient’s healthy limb can be used for comparison with the affected side in the presence of orthopaedic impairments \citep{Ankle_normal}. The principal objective of this study is to assess the recovery of orthopaedic patients and compare their gait with that of a healthy person. 

\subsection{Cataract Surgery Data}\label{Cat_data}
\begin{figure}[!htb]
	\centering
	\includegraphics[width=1.0\textwidth]{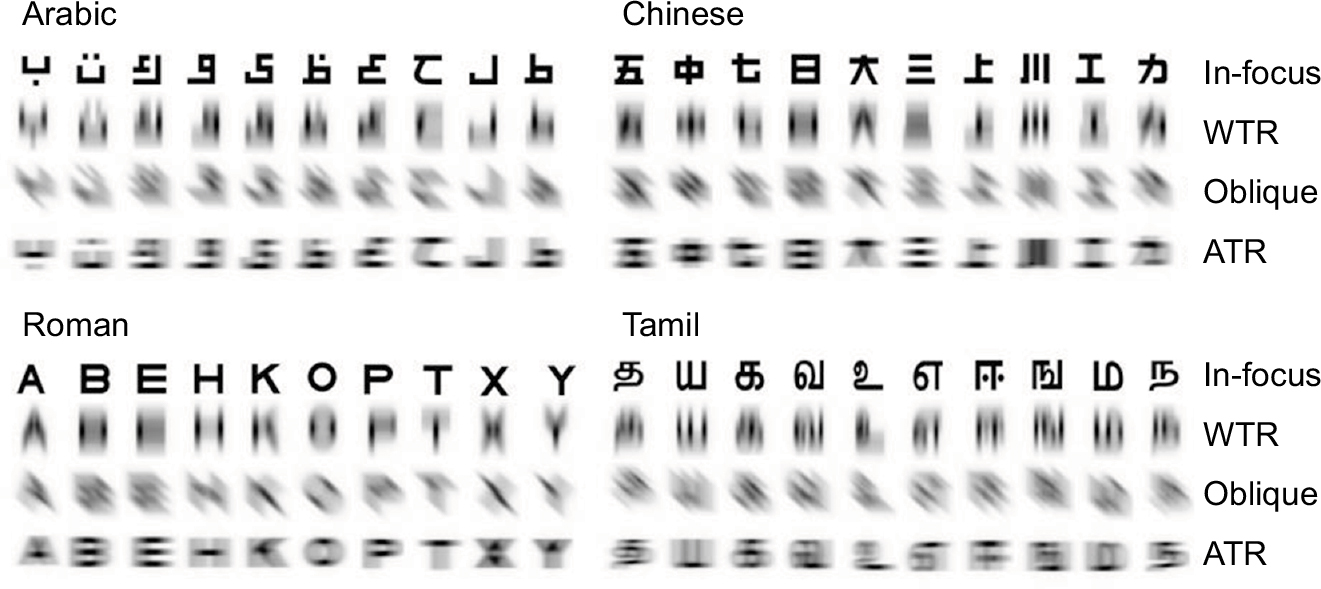}
	\caption{Visual distortions due to ATR, WTR and Oblique astigmatism.}
	\label{fig:illus2}
\end{figure}

A cataract is a clouding that develops in the natural lens of the eye or its envelope. In due course of time, the lens loses its transparency and leads to partial or total loss of vision. This has been documented to be the most significant cause of bilateral blindness in India \citep{Catind, catind1, Catind2}.
Cataract surgery is the removal of the opaque natural lens from the eye, and an artificial intra-ocular lens implant is then inserted to restore vision. India is a signatory to the World Health Organization resolution on VISION 2020: The Right to Sight. Efforts from all stakeholders have resulted in an increased number of cataract surgeries performed in India \citep{WHO}. It is well-known that one common side effect of the cataract surgery is that the incision causes unwanted changes to the natural corneal shape causing an astigmatic eye. The refractive error of the astigmatic eye induces several focal points in different directions. For example, the image may be perfectly focused on the retina in the horizontal (sagittal) plane, but not in the vertical (tangential) plane. There are two types of astigmatism based on the axes of the principal meridians - regular or irregular. In this paper, we consider regular astigmatism, where the principal meridians are perpendicular. For regular astigmatism, its axis ranges from $\ang{0}$ to $\ang{180}$. In general, regular astigmatism can be subdivided into three types: (i) With-the-rule astigmatism (WTR) - the vertical meridian ($\ang{90}$) is steepest (e.g. an American football lying on its side), (ii) Against-the-rule astigmatism (ATR)- the horizontal meridian ($\ang{180}$) is steepest (e.g. an American football standing on its end), and (iii) Oblique astigmatism - the steepest curve lies in $(\ang{120}, \ang{150})$ or $(\ang{30}, \ang{60})$ \citep{morlet2001, mimoni2013}.
The eye affected by WTR astigmatism sees vertical lines more sharply than horizontal lines. The situation is reversed for ATR astigmatism. Oblique astigmatism is worse than WTR and ATR astigmatism because most of the standard objects (e.g. letters) in our surroundings are horizontal or vertical. The objects get distorted horizontally or vertically for WTR and ATR astigmatism, whereas the distortion is severe for oblique astigmatism. The visual distortions for different types of astigmatism are displayed in Figure \ref{fig:illus2}. It is of primary interest for ophthalmologists to study post-operative astigmatism and visual recovery over time \citep{Recover}. In this article, we consider a study conducted at Disha Eye Hospital and Research Center, Barrackpore, West Bengal, India, over a period of two years (2008-10). In total, 54 patients were operated, and the axes of astigmatism were measured on the 1st, 7th, and 15th days after the surgery. See \citet{bakshi10} for a detailed description of the study and data description. The main objective of regular monitoring is to study the process of visual recovery and identify patients who require additional care to minimize post-operative trauma. In particular, medical practitioners are interested in foreseeing the improvement of patients based on previous inspections.

\subsection{The challenge in analysing Cataract Surgery and Abnormal Gait Data}\label{Challenge}
\begin{figure}[!htb]
	\centering
	\includegraphics[width=0.6\textwidth]{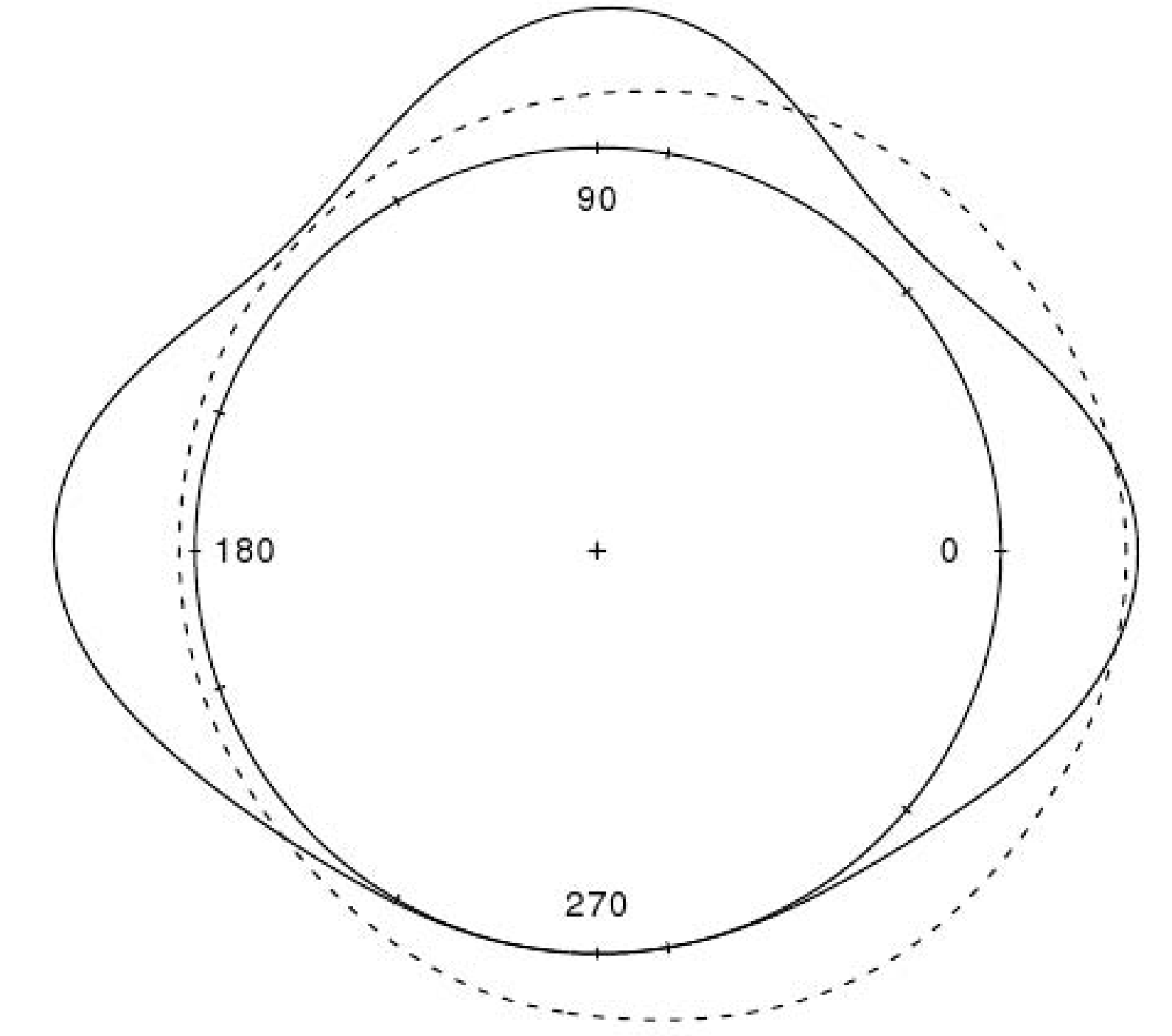}
	\caption{Density plots of a multimodal circular distribution (solid lines) and its transformed unimodal distribution (dotted lines) obtained by multiplying with 4.}
	\label{fig:illus}
\end{figure}

\begin{figure}
	\centering
	\includegraphics[width=0.5\textwidth]{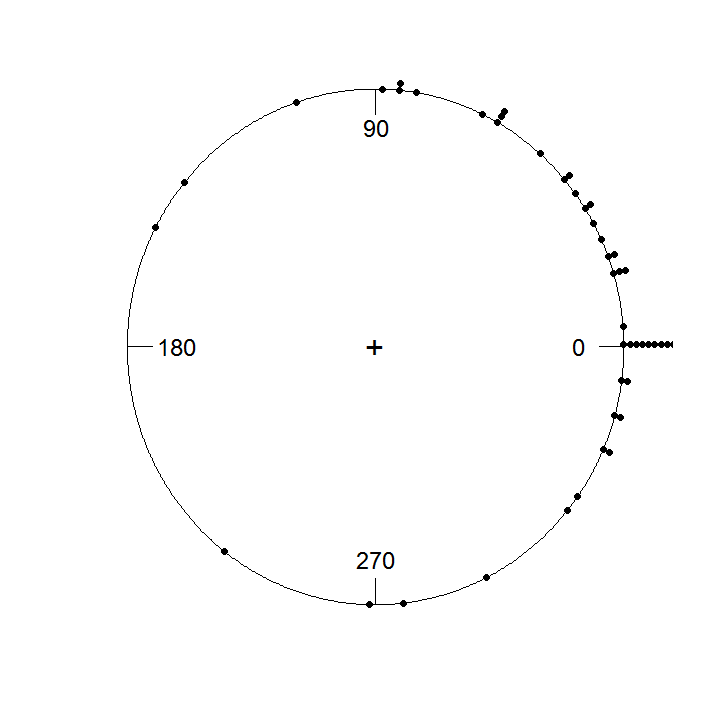}\includegraphics[width=0.5\textwidth]{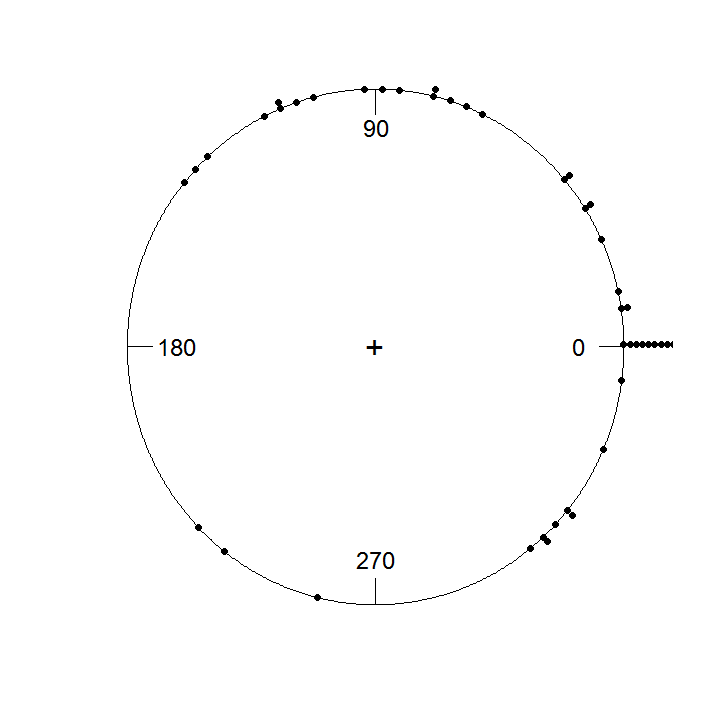}
	\caption{Circular plots of the transformed axes of astigmatism after 15 days (left) and after 7 days (right).}
	\label{cirplo}
\end{figure}

There are some critical issues involved in the modeling of astigmatism data under consideration. According to medical practitioners, if the axis is closer to $\ang{0}$, $\ang{90}$ or $\ang{180}$, then it is not a matter of serious concern as most of the standard objects in nature are horizontal or vertical. Therefore, it is preferred that the axis is closer to $\ang{0}$, $\ang{90}$ or $\ang{180}$. In order to make only one preferred direction, the observed angles are multiplied with 4 and then transformed by taking mod $\ang{360}$. Consequently, the preferred angle reduces to $\ang{0}$ ( $=\ang{360}$) and hence, the multimodal distribution becomes a distribution having a single mode at $\ang{0}$. The effect of this transformation on circular distribution is illustrated in Figure \ref{fig:illus}.
%The original and the transformed distributions are presented in  Figure \ref{fig:illus}. 
In this particular study, the measurements were taken up to the precision of $\ang{1}$. Therefore, the transformed variables become zero-inflated due to the high concentration of observations censored in the interval $(-\ang{2}, \ang{2})$.
%Therefore, due to this transformation and rounding, the variables become zero-inflated with precision of $\ang{4}$. 
The transformed axes of astigmatism after 7 days and 15 days of the surgery are presented using circular plots in Figure \ref{cirplo}. A similar issue arises related to the measurements on ankle extension in abnormal gait data. The measurements were taken up to the precision of $\ang{0.5}$ and averaged over $20$ strides of each individual. Almost $50\%$ of the observations on healthy individuals, and $60\%$ of the observations on orthopaedically impaired patients, are recorded as $\ang{0}$.


Numerous studies focusing on zero-inflated random variables in the linear setup are available in the literature.  \citet{tobin582436}, \citeauthor{heckman74679694} (\citeyear{heckman74679694}, \citeyear{heckman79127145}) proposed some models in the context of linear regression where the responses are zero-inflated. \citet{lambert92114} considered Poisson regression for count data with excess zeros in the response variable. \citet{bhuyan16tr} discussed the case when both the responses and covariates are zero-inflated and proposed estimation methodology under a Bayesian setup. See \citet{min02733} for a detailed review of the zero-inflated regression models. However, in all of these works, the response and the covariate are linear in nature, where zero-inflation is caused by censorship through a selection mechanism. The case of circular random variables is considerably different from linear ones. The difference mainly arises due to the topology of the circle, where the zero cannot be considered to be located at the boundary of the sample space. In this context, a spike at zero may be interpreted in the same way as a spike at any other angle, since the origin can be fixed arbitrarily without loss of generality.
There are only a few works on the analysis of zero-inflated circular data in the literature. The modeling of a circular random variable with point-accumulation was studied in \citet{Biswas16}. In the context of circular-circular regression, \citet{Jhatr1705} proposed a model with a zero-inflated response variable and discussed the associated inferential issues. However, there is no model available in the literature for the case of a zero-inflated circular covariate. To avoid such difficulty, \citet{Jhatr1705} analysed a subset of the dataset, considering only non-zero covariate values. It is important to note that the conventional circular-circular regression model and the model proposed by \citet{Jhatr1705} are not appropriate for handling excess zeros in the covariate, and provide biased results. Moreover, the existing models are not capable of joint modeling of the periodical observations on astigmatism recorded over three different inspections. To model such data, we propose a two-stage circular-circular regression model based on continuous latent variables.
In contrast to the assumption of randomly occurring zeros in \citet{Jhatr1705}, we consider a more realistic assumption that zero inflation occurs due to censoring. In Section \ref{method}, we discuss the modeling approach and propose an estimation methodology under a Bayesian setup using the Markov chain Monte Carlo (MCMC) algorithm. Some generalisations and a special case of the proposed model are also discussed in the same section. The performance of the proposed method is compared with the existing competitors through simulation in Section \ref{sim}. Analysis of real datasets on post-operative astigmatism and abnormal gait are presented in Section \ref{Case}. The key findings are summarised and concluded with some discussions on future research in Section \ref{conc}.

%%which is useful in terms of interpretation when there is a preferred direction for the observations as in our data and we propose a method to compare the recovery over two weeks of the surgery

\section{Proposed Model and Methodology}{\label{method}}

\begin{figure}
	\centering
	%\begin{tabular}{cc}
	\includegraphics[scale=0.4]{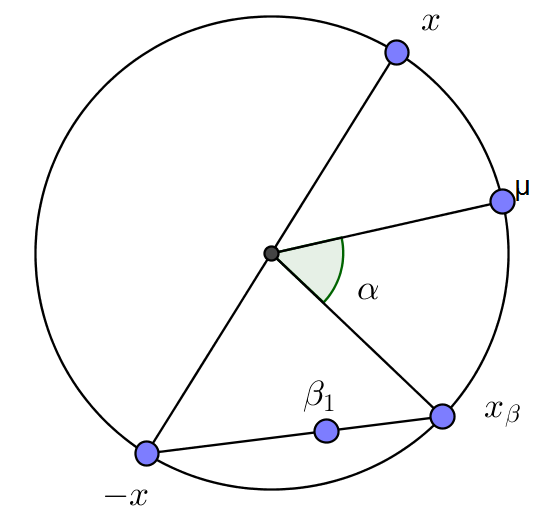} %\includegraphics[width=0.6\textwidth]{ccregression.png}
	%\end{tabular}
	\caption{Circular-circular regression model.}
	\label{fig:frog}
\end{figure}
In this section, we first describe the M\"obius transformation based circular-circular regression model proposed by \citet{Kato08}, which is a reparameterization of the model proposed by \citet{downs02683697}.
This reparameterization induces a nice geometry to the regression link function. When the angular error follows wrapped Cauchy distribution, it provides some advantages in terms of distributional properties as the wrapped Cauchy distribution is closed under rotation and M\"obius transformation. See \citet{Kato08} for details. It is important to note that the rotational model proposed by \citet{mackenzie576162}, where the predicted mean direction of the response is a fixed rotation of the covariate, is a special case of the M\"obius transformation based model. Unlike the rotational model, this model is also appropriate when there is a high concentration of observed responses on a section of the unit circle. This unique feature makes the model parameters easily interpretable and provides interesting insights.

Let us represent circular random variables $\theta_Y$ and $\theta_X$ as complex random variables $Y= e^{i \theta_Y}$ and $X= e^{i \theta_X}$, respectively taking values on the circumference of a unit circle. The circular-circular regression model of \citet{Kato08} is represented as:

\be
Y=\beta_0 \frac{X+\beta_1}{1+\bar{\beta_1}X}\epsilon,
\label{kat}
\ee
where $\beta_0$, $\epsilon$ $\in \{z:z \in \mathbbm{C};|z|=1\}$, $\beta_1 \in \mathbbm{C}$, and the angular error $\arg(\epsilon)$ follows a wrapped Cauchy distribution with mean direction $0$.  
%However, other distributions for angular error, %such as the von Mises distribution and wrapped %Normal distribution, can also be considered. 
The regression link function, which is a form of M\"obius transformation, is a mapping from a unit circle onto itself. Here, $\beta_0$ is the rotation parameter because the multiplication by a unit complex number is an anti-clockwise rotation by the argument of the same unit complex number. In this case, the predicted mean $\mu(\theta_x)$ given $x$ is obtained by rotating $x_\beta$ by $\alpha=\arg(\beta_0)$, where $x_\beta=\frac{x+\beta_1}{1+\overline{\beta_1}x}$ is the intersection of the unit circle with the line joining $-x$ and $\beta_1$ (see Figure \ref{fig:frog}). For $|\beta_1|>1$, the regression link function can be  geometrically represented as a straight line connecting $1/\overline{\beta_1}$ and $\frac{\beta_1}{|\beta_1|}\frac{\beta_1}{|\beta_1|}\overline{x}$. Therefore, it also covers the cases in which the predicted mean direction depends on the conjugate of $x$ (i.e. when there is a reflection of $x$). The intersection of this line with the unit circle is then rotated by $\alpha$ to obtain the predicted mean direction $\mu$. If $|\beta_1|$ is closer to 1 and $x$ is uniformly distributed, then $x_\beta$ is highly concentrated around $\frac{\beta_1}{|\beta_1|}$. The distribution of $Y$ becomes independent of $x$ if $|\beta_1|=1$. For $|\beta_1|=0$, the predicted mean direction is just a rotation of $x$, i.e. $\mu=\beta_0 x$. See \citet{Kato08} for more details.

To model the axis of astigmatism recursively based on three consecutive inspections, we extend the circular-circular regression model (\ref{kat}) in a  two-stage setup as:
\be
Y=\beta_0 \frac{X+\beta_1}{1+\overline{\beta_1} X}\epsilon_1,
\label{reg12}
\ee
\be
X=b_0 \frac{V+b_1}{1+\overline{b_1} V}\epsilon_2,
\label{reg22}
\ee
where $\beta_0$, $b_0$, $\epsilon_1$, $\epsilon_2$  $\in \{z:z \in \mathbbm{C};|z|=1\}$; $\beta_1$ ,$b_1$ $\in \mathbbm{C}$, $Y=e^{i\theta_{Y}}$, $X=e^{i\theta_{X}}$ and $V=e^{i\theta_{V}}$. We assume $\arg(\epsilon_i) \sim WC(0,\rho_i)$ for $i=1,2$, where $WC(\mu,\rho)$ represents the wrapped Cauchy distribution with parameters $\mu \in [0,2\pi)$ and $\rho \in [0,1]$. Also, $\arg(\epsilon_1)$ and $\arg(\epsilon_2)$ are assumed to be independently distributed. In the literature of Econometrics, $V$ is known as instrumental variable \citep[p-97]{Inst}, which may not directly effect the response $Y$ but it can induce changes only through the covariate $X$. 
%In the usual two-stage regression framework in, one first fits the model given by equation (\ref{reg22}), and then fits the model given by equation (\ref{reg12}) replacing $X$ with its predicted values. %\textcolor{red}{Unlike the usual framework in frequentist setting, the purpose of the two-stage regression modelling in our paper is handling the zero-inflation in the covariate as well as the response. Therefore, we consider the substitution of $X$ only when it is zero by its predicted value in the censored interval based on equation (\ref{reg22}).} 
This modeling approach allows us to compare the visual recovery of the patients over two consecutive weeks.

Note that the circular-circular regression model,  given by  equation (\ref{kat}), is inadequate for handling zero-inflated data \citep{Jhatr1705}. Similarly, the aforementioned two-stage model will give very poor fit to the data due to point accumulation at zero for both the response and the covariate.
In order to model zero-inflated circular data, we first define circular latent variables $\theta_{Y^*}$ and $\theta_{X^*}$ as
$$\theta_{Y} = \begin{cases} 0, &  \mbox{if } \theta_{Y^*} \in (-\delta_Y,\delta_Y), \\ \theta_{Y^*}, & \mbox{otherwise,} \end{cases}$$
and
$$\theta_{X} = \begin{cases} 0, & \mbox{if } \theta_{X^*} \in (-\delta_X,\delta_X), \\ \theta_{X^*}, & \mbox{otherwise,} \end{cases}$$
respectively, where $\delta_Y$, $\delta_X$ are constants taking values in $[0,\pi)$. Now we propose the two-stage circular-circular regression model based on the aforementioned latent variables as
\be
Y^*=\beta_0 \frac{X^{\ast}+\beta_1}{1+\overline{\beta_1} X^*}\epsilon_1,
\label{reg1}
\ee
\be
X^*=b_0 \frac{V+b_1}{1+\overline{b_1} V}\epsilon_2,
\label{reg2}
\ee
where
$\beta_1$, $b_1$ $\in \mathbbm{C}$,
$Y^*=e^{i\theta_{Y^*}}$, $X^*=e^{i\theta_{X^*}}$.
In the absence of instrumental variable, one can still use (\ref{reg1}) and (\ref{reg2}) for modeling of zero-inflated response and covariate with $|b_{1}|=1$.
Note that the above model reduces to the two-stage circular-circular regression model without zero-inflation, given by (\ref{reg12}) and (\ref{reg22}), when $\delta_X=\delta_Y=0$.

\subsection{Bayesian Estimation}\label{estimation}
We propose a Bayesian estimation of the model parameters involved in the two-stage model given by \eqref{reg1} and \eqref{reg2}, using the MCMC algorithm based on data augmentation. In the conventional frequentist approach, computational challenges arise in the model fitting due to intractable numerical integration involved in the log-likelihood function. Unlike the usual frequentist set-up, the proposed Bayesian approach provides simultaneous estimation of the model parameters and a natural framework for prediction over unobserved data. Thus, by generating posterior predictive densities, rather than point estimates, we can make probability statements giving greater flexibility in presenting results. For instance, we can discuss findings concerning specific hypotheses or in terms of credible intervals which can offer a more intuitive understanding for the practitioners.

Let us denote the observed data by $D=\{(\theta_{V_{i}},\theta_{X_{i}},\theta_{Y_{i}}), i=1\ldots n\}$, and the parameter vector by $\Theta_i=(\theta_{0i},r_i,\theta_{1i},\rho_i)$ for $i=1,2$,  where $\beta_0=e^{i\theta_{01}}$, $\beta_1=r_1e^{i\theta_{11}}$, $b_{0}=e^{i\theta_{02}}$, $b_{1}=r_2e^{i \theta_{12}}$,  $r_1, r_2 \in [0,\infty)$. Also, denote the density of the wrapped Cauchy distribution and the truncated wrapped Cauchy distribution by $f_{W}(\theta;\mu,\rho)$ and $f_{TW}(\theta;\mu,\rho,-\delta,\delta) =K^{-1} f_{W}(\theta;\mu,\rho) \mathbbm{1}(\theta \in (-\delta,\delta))$, respectively, where $K=\int_{-\delta}^{\delta}f_{W}(\theta;\mu,\rho)d\theta$. The joint posterior density of the model parameters and the latent variable involved in equation (\ref{reg1}) is given by

\begin{align*}
\pi(\Theta_1,\theta_{Y^*}|D) \propto \pi(\Theta_1) \times \prod_{i=1}^{n} \left[f_{W}(\theta_{Y_i};\mu_{1i},\rho_1)\mathbbm{1}(\theta_{Y_i}\neq 0) \right. \\ \left. 
+ K f_{TW}(\theta_{Y_i^*};\mu_{1i},\theta_{X_i^*},-\delta_Y,\delta_Y)\mathbbm{1}(\theta_{Y_i}=0)\right]\nonumber, 
\end{align*}
where $\mu_{1i}=\arg(\beta_0\frac{X_i^*+\beta_1}{1+\overline{\beta_1} X_i*})$, and $\pi(\Theta_1)$ denotes the prior density of $\Theta_1$. Similarly, the joint posterior density of the model parameters and the latent variable involved in equation (\ref{reg2}), is given by
\begin{align*}
\pi(\Theta_2,\theta_{X^*}|D) \propto \pi(\Theta_2) \times \prod_{i=1}^{n} \left[f_{W}(\theta_{X_i};\mu_{i2},\rho_2)\mathbbm{1}(\theta_{X_i}\neq 0) \right. \\ \left.   + K f_{TW}(\theta_{X_i^*};\mu_{i2},\rho_2,-\delta_X,\delta_X)\mathbbm{1}(\theta_{X_i}=0)\right] \nonumber, 
\end{align*}
where $\mu_{2i}=\arg(b_0\frac{V_{i}+b_1}{1+\overline{b_1} V_{i}})$, and $\pi(\Theta_2)$ denotes the prior density of $\Theta_2$. The full conditional densities of the latent variables $\theta_{X_i^*}$ and $\theta_{Y_i^*}$ have the following closed-form expressions

\be
\pi(\theta_{X_i^*}|\Theta_1,\Theta_2,D) \equiv  \begin{cases} \mathbbm{1}(\theta_{X_i^*}=\theta_{X_i}),  & \mbox{if } \theta_{X_i} \neq 0 \\ f_{TW}(\theta_{X_i*};\mu_{2i},\rho_2,-\delta_X,\delta_X), & \mbox{otherwise,} \end{cases}
\label{tw1}
\ee

and

\be 
\pi(\theta_{Y_i^*}|\Theta_1,\Theta_2,D) \equiv \begin{cases} \mathbbm{1}(\theta_{Y_i^*}=\theta_{Y_i}), & \mbox{if } \theta_{Y_i} \neq 0 \\ f_{TW}(\theta_{Y_i^*};\mu_{1i},\rho_1,-\delta_Y,\delta_Y), & \mbox{otherwise,} \end{cases}
\label{tw2}
\ee

respectively. We propose an algorithm for generating samples from the truncated wrapped Cauchy distribution which is discussed in Subsection \ref{SA}. The full conditional densities of the model parameters $\Theta_1$ and $\Theta_2$ are given by 
\be 
\pi(\Theta_1|\Theta_2,\theta_{X^*},\theta_{Y^*},D) \propto \pi(\Theta_1)\prod_{i=1}^{n} f_{W}(\theta_{Y_i^*};\mu_{1i},\rho_1),
\label{fc11}
\ee

\be 
\pi(\Theta_2|\Theta_1,\theta_{X^*},\theta_{Y^*},D) \propto \pi(\Theta_2)\prod_{i=1}^{n} f_{W}(\theta_{X_i^*};\mu_{2i},\rho_2),
\label{fc12}
\ee
respectively. Note that the full conditionals of $\Theta_1$ and $\Theta_2$ cannot be expressed in closed form. Therefore, we employ the Metropolis-Hastings algorithm for generating samples from the posterior densities of the parameters and the detailed algorithm is provided in Subsection \ref{SA}. In our context, $\delta_Y$ and $\delta_{X}$ are known a priori. However, in many situations, these are not known and one can consider a suitable prior as discussed in Subsection \ref{SA}.

\subsection{Sampling Algorithms}\label{SA}

\subsubsection{Sample Generation from Truncated Wrapped Cauchy Distribution}\label{Truncated}
Let $\theta_Z$ be a circular random variable following truncated wrapped Cauchy distribution with pdf $f_{TW}(\theta_z;\mu,\rho,a,b)$, where $a,b \in [-\pi,\pi)$. The support of $f_{TW}(\cdot)$ is $(a,b)$, if $a<b$, and $(a,\pi) \cup [-\pi,b)$, otherwise. To generate samples from $f_{TW}(\theta_z;\mu,\rho,a,b)$, for $a<b$, we simulate observations from $f_{W}(\theta_z;\mu,\rho)$ and accept the observations which lie in $(a,b)$. Similarly, for $a>b$, we accept the observations lying in $(a,\pi) \cup [-\pi,b)$. However, the acceptance rate is very low for small values of $(b-a)1(a<b)+\left[2\pi-(b-a) \right] {1(a>b)}$. For example, the acceptance rate is approximately $0.03\%$ when $a=\pi-0.035$, $b=-\pi+0.035$, $\mu=0$, and $\rho=0.95$. Thus, we propose a novel algorithm for generating samples from the truncated wrapped Cauchy distribution based on the geometry of the M\"obius transformation.

\begin{figure}
	\centering
	%\begin{tabular}{cc}
	\includegraphics[width=0.6\textwidth]{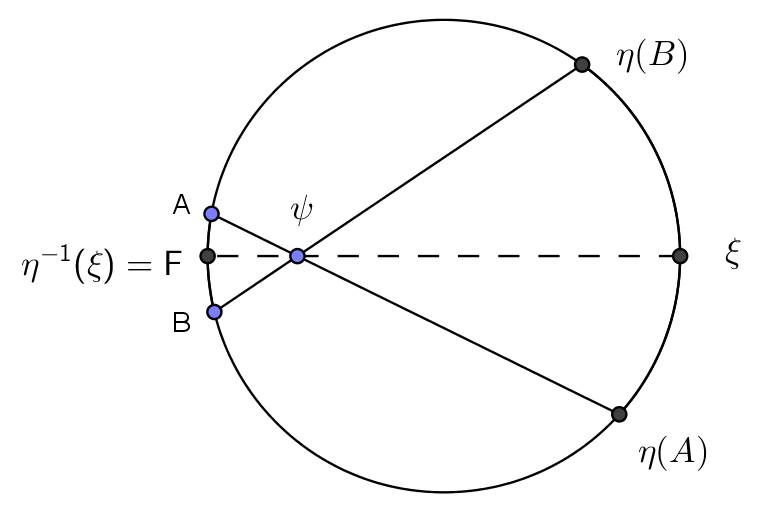} %\includegraphics[width=0.6\textwidth]{ccregression.png}
	%\end{tabular}
	\caption{Sample Generation from truncated wrapped Cauchy distribution.}
	\label{fig2}
\end{figure}

If $\theta_Z$ is uniformly distributed in $[0,2\pi)$ and $Z=e^{i\theta_z}$, then $\arg\left(\frac{\psi-Z}{1-\overline{\psi} Z}\right)$ follows wrapped Cauchy distribution with parameters $\mu=\arg(\psi)$, and $\rho=|\psi|$, where $\psi \in \{c \in \mathbbm{C}:|c| \leq 1\}$ \citep{Kato08}. The geometry of the transformation $\eta(Z)=\frac{\psi-Z}{1-\overline{\psi}Z}$ is presented in Figure \ref{fig:frog} with $x=-Z$, $\beta_1=\psi$ and $\alpha=0$. Note that, $\eta(Z)$ is the point on the circumference of the unit circle which is situated at the intersection of the line joining $Z$ and $\psi$ and the unit circle. It is easy to see that $\arg\{\eta(Z)\}$ follows wrapped Cauchy distribution with $\mu=\arg(\psi)$ and $\rho=|\psi|$. To simulate an observation $U$ from the truncated wrapped Cauchy distribution, first generate a unit complex number $\xi$ uniformly in the region between $\eta(A)$ and $\eta(B)$ where $A=e^{ia}$ and $B=e^{ib}$. Then, consider the argument of its inverse M\"obius transformation $\eta^{-1}(\xi)$ as $U$. This construction is diagrammatically illustrated in Figure \ref{fig2} and summarized in a  simple algorithm below.\\

\textbf{Algorithm 1: Sampling from Truncated Wrapped Cauchy Distribution}
\begin{itemize}
	\item[Step 1:] Take $A=e^{ia}$, $B=e^{ib}$ and choose a point $c$ in the support of $f_{TW}(\theta_z;\mu,\rho,a,b)$.
	
	\item[Step 2:] Generate a random unit complex number $\xi$ uniformly  from the arc joining $\eta(A)$ and $\eta(B)$ containing $\eta(C)$, where $C=e^{ic}$.

	\item[Step 3:] Take $U=\arg\{(\eta^{-1}(\xi)\}$.
\end{itemize}

\subsubsection{Metropolis-Hastings Algorithm}\label{MH}
For the purpose of Bayesian estimation, we consider the following prior distributions for the parameter vectors $\Theta_1$ and $\Theta_2$. The joint prior distribution $\pi(\Theta_i)$  can be expressed as the product of $\pi(\theta_{0i}) \propto 1$, $\pi(\theta_{1i}) \propto 1$, $\pi(r_i) \propto e^{-r_i^2}$ and $\pi(\rho_i) \propto \rho_i^{a_{\rho_i}-1}(1-\rho_i)^{a_{\rho_i}-1}$, for $i=1,2$. It can be easily verified that  the joint posterior density is proper for $a_{\rho_i}>1$. Similar priors have been considered by \citet{ravindran11547561} in the context of circular-circular and circular-linear regressions. For the purpose of implementing  Metropolis-Hastings Algorithm, we consider the proposal distributions for $\theta_{01},\theta_{11},\theta_{02},\theta_{12},\rho_1,\rho_2$ to be uniform, and the proposal distributions for $r_1,r_2$ are chosen to be exponential. As discussed before, $\delta_Y$ and $\delta_{X}$ may be unknown and one can employ the following MCMC algorithm with priors $\pi(\delta_{Y}) \propto 1$ and $\pi(\delta_{X}) \propto 1$, and uniform proposal densities.\\

\textbf{Algorithm 2: Metropolis Hastings Algorithm}\\
\begin{itemize}
	\item[Step 1:] Sample $\theta_{X^*}$ from the density $\pi(\theta_{X^*}|\Theta_1,\Theta_2,D)$ given by equation \eqref{tw1}, and then sample $\theta_{Y*}$ from the density $\pi(\theta_{Y^*}|\Theta_1,\Theta_2,D)$ given by equation \eqref{tw2}. 
	
	\item[Step 2:] Generate the model parameters sequentially from the corresponding proposal densities and denote it as $\nu_p'$. Given the previous value of $\nu_p$ and the current
	draw $\nu_p'$, return $\nu_p'$ with probability
	$$\alpha_{MH}(\nu_p,\nu_p')=\min\left\{1,\frac{\pi(\nu_p'|-)\pi(\nu_p',\nu_p)}
	{\pi(\nu_p|-)\pi(\nu_p,\nu_p')}\right\},$$
	where $\pi(d,w)$ denotes the proposal density at $d$ with parameter $w$, and $\pi(\cdot|)$ denotes the full conditional densities given by equations \eqref{fc11} and \eqref{fc12}.
	Otherwise, repeat the previous value $\nu_p$.
	
	\item[Step 3:] Repeat Step 1 and Step 2 until convergence.
\end{itemize}

\subsection{Special case } \label{redmo}
Under certain restrictions on the model parameters, the geometry of the M\"obius transformation based link function provides interesting insights when there is a preferred direction for the response variable.  Without loss of generality, we consider $\ang{0}$ to be the preferred direction. Then, we consider a special case of the circular-circular regression model given by equation (\ref{kat}), where  $\arg(\beta_0)=\arg(\beta_1)=0$ and $\beta_1 \in [-1,1]$. Under this restriction, the symmetry about the preferred direction for the predicted responses and the corresponding covariate values is maintained as $\arg\{\mu(\theta_x)\}=-\arg\{\mu(-\theta_x)\}$. This can be seen directly from the geometry as shown in Figure \ref{fig:frog} by taking $\beta_1 \in [-1,1]$ and $\alpha=0$.  In this case, the larger the value of $\beta_1$, the more is the tendency of the predicted response to move towards $\ang{0}$. The constraint $\beta_1 \in [-1,1]$ ensures that both the predicted mean direction and the corresponding covariate value lie on the same semi-circle having the diameter as the line segment joining $\ang{0}$ and $\ang{180}$.
This implies that the predicted mean direction lies on the shortest arc joining the respective covariate value and $\ang{0}$ when $\beta_1>0$. This special case also ensures that $\mu(\theta_{x_1})<\mu(\theta_{x_2})$, for all $\theta_{x_1}<\theta_{x_2}$, where $\theta_{x_1},\theta_{x_2} \in [0,\pi)$. That is, the closer the covariate value is to $\ang{0}$, the closer is the corresponding predicted mean response to $\ang{0}$.

%This phenomenon occurs in the cataract surgery data as the closer the observation is to $\ang{0}$, the healthier is the eye.  

This special case can be reparameterized and presented as the centered regression model proposed by \citet{downs02683697} with both the angular location parameters fixed at 0. The corresponding regression link function can be written as
$\arg\{\mu(\theta_x)\}=2 \arctan \{\omega \tan (\frac{\theta_x}{2})\}$, where $\omega \in [0,\infty)$. As the angular parameters are fixed at 0, there is a one-one monotonic correspondence between $\omega$ and $\beta_1$, given by $\beta_1=\frac{1-\omega}{1+\omega}$. Therefore, the parameter $\omega$, if greater than or less than 1, indicates the general tendency of the predicted response to move towards or away from $\ang{0}$ as compared to its covariate value, respectively.

%%	We next fix $\beta_1 \in [-1,1]$, which ensures that both the covariate and the predicted mean direction lie on the same semicircle if we consider the two semi-circles having the diameter as the line joining $\ang{0}$ and $\ang{180}$. 
%%	Then, $\theta_x \in [0,\pi) \implies \arg\{\mu(\theta_x)\} \in [0,\pi)$ and  $\theta_x\in [-\pi,0) \implies \arg\{\mu(\theta_x)\} \in [-\pi,0)$.  
%% This holds true for both the stages combined also.

For the two-stage circular-circular regression model given by equations (\ref{reg1}) and (\ref{reg2}), we consider  $\arg(b_0)=\arg(b_1)=\arg(\beta_0)=\arg(\beta_1)=0$, and $b_1,\beta_1 \in [-1,1]$. As compared to the unrestricted model, this special case enforces all the predicted response to either move towards or away from zero as compared to its respective covariate value.  Therefore, the two stages of the regression model can be compared in terms of their tendency to move towards $\ang{0}$ on the basis of the posterior distributions of $\beta_1$ and $b_1$. In contrast to the general model, we  consider uniform prior and proposal distributions for $r_1$ and $r_2$. The choices of prior and proposal distributions for all the other parameters are considered same as discussed in Subsection \ref{MH}. When there is a preferred direction, it is easier to compare two regression models based on the estimates obtained from the special case as compared to Model I. This special case is parsimonious and easy to interpret. However, a drawback of this special case as compared to the unrestricted model is that all the predicted mean responses simultaneously move towards or away from the preferred direction. Therefore, this model is not appropriate for prediction when some of the observed responses are closer, and some are farther from the preferred direction compared to their respective covariate values. We consider this special case for data analysis to provide some insights into the cataract surgery data based on the posterior distributions of the model parameters.

\subsection{Some Generalisations}\label{gen}
The proposed two-stage model, given by (\ref{reg1}) and (\ref{reg2}), can be extended to the case when there are multiple circular covariates. \citet{Jha17sm} proposed a multiple circular-circular regression model (MCR2) based on M\"obius transformation which is represented as:

\be
Y=\beta_0 \frac{X^{(s)}+\beta_1}{1+\overline{\beta_1}X^{(s)}}\epsilon,
\label{mcr}
\ee
where $X^{(s)}=\frac{\sum_{j=1}^{k}p_jA_jX_j}{|\sum_{j=1}^{k}p_jA_jX_j|}$,
$A_1=1$,
$A_j \in \{z:z \in \mathbb{C};|z|=1\}$ for $j=2,\ldots k,$  $p_j \in [0,1]$ for $j=1,\ldots k,$ $\sum_{j=1}^{k}p_{j}=1$,
and $X_{1}, \ldots, X_{k}$ are covariates.
%$;0\leq p_j \leq 1~\mbox{ for } ~j=1,\ldots ,k\sum_{j=1}^{k}p_{j}=1.$ 
Note that the model (\ref{mcr}) reduces to (\ref{kat}) for $k=1$. Now, without loss of generality, we consider that $X_{1}$ is a zero-inflated covariate and define a latent circular variable $X^{*}_{1}$ as:
$$\theta_{X_{1}} = \begin{cases} 0, & \mbox{if } \theta_{X^{*}_{1}} \in \left(-\delta_{X_{1}},\delta_{X_{1}} \right),\\
\theta_{X^{*}_{1}}, & \mbox{otherwise,} 
\end{cases} $$
where $\theta_{X_1^{\ast}}=\arg(X_1^{\ast})$.
Note that the the response $Y$ is also zero-inflated and we consider the latent response $Y^{\ast}$ as defined in Section \ref{method}.
Then, the generalised two-stage circular-circular regression model for zero-inflated data with multiple covariates is given as:
$$Y^{*}=\beta_0 \frac{X^{(a)}+\beta_1}{1+\overline{\beta_1}X^{(a)}}\epsilon_1,$$
and
$$X_1^*=b_0\frac{W+b_1}{1+\overline{b_1}W}\epsilon_2,$$
where $X^{(a)}=\frac{p_1A_1X_1^* +\sum_{j=2}^{k}p_jA_jX_j}{|p_1A_1X_1^* +\sum_{j=2}^{k}p_jA_jX_j|}$, $W=\frac{\sum_{i=1}^{l}q_i B_i W_i}{|\sum_{i=1}^{l}q_iB_iW_i|}$,
$B_1=1$,
$B_i \in \{z:z \in \mathbb{C};|z|=1\}$ for $i=2,\ldots l$, $q_i \in [0,1]$, $\sum_{i=1}^{l}q_{i}=1$,
and $W_{1},\ldots, W_{l}$ are covariates used to regress $X_{1}^{\ast}$. For the purpose of Bayesian estimation, one can consider Dirichlet priors for $(p_1,\ldots p_k)$ and $(q_1,\ldots q_l)$ with parameter vectors $\textbf{1}_{k\times 1}$ and $\textbf{1}_{l \times 1}$, respectively. The priors for $\arg(A_j)$, for $j=2,\ldots,k$, and $\arg(B_i)$ for $i=2,\ldots,l$, can be taken as uniform. The MCMC algorithm mentioned in Subsection \ref{SA} is readily extended to this generalised model. Further, one can generalise the proposed model for modeling data with multiple points of accumulation due to asymmetric censoring intervals.

\section{Simulation Studies}\label{sim}
In order to study the performance of the proposed method, we generate data considering different choices of parameter values for two different sample sizes, 50 and 100. We generate $\theta_{V}$ from von Mises distribution with mean 0 and concentration parameter $2$. The five different sets of parameters values of $\Theta_{1}$ and $\Theta_{2}$ are chosen, keeping $\delta_{X}=\delta_{Y}=0.035$ radians ($\ang{2}$), such that the approximate proportions of zeros in the response and covariate are given by $(0.15,0.15)$, $(0.10,0.10)$, $(0.10,0)$, $(0,0.10)$ and $(0,0)$, respectively. We consider the priors as discussed in the Subsection \ref{SA} with $a_{\rho_i}=2$ for $i=1,2$. 
We generate 100,000 samples from the posterior distributions of the associated model parameters using the MCMC algorithm and find the posterior mean and standard deviation of the linear parameters based on every 10th iterate discarding the first 40,000 iterations as burn-in. Similarly, we compute the posterior circular mean and circular dispersion for the circular parameters. This is repeated 500 times and the average estimates are reported in Tables \ref{t1}-\ref{t5}. Note that, the circular dispersion (c.d.) for $n$ circular observations $\phi_1,\ldots \phi_n$ is given by $1-\bar{R}$, where $\bar{R}=\frac{||\sum_{i=1}^{n} z_i||}{n}$ and $z_i=(\cos \phi_i,\sin \phi_i)$ for each $i=1,\ldots, n$. We also report the coverage probability (CP) corresponding to all the parameters. As expected, the standard deviation (s.d.) and c.d. decrease as the sample size increases (See Tables \ref{t1}-\ref{t5}). In this limited simulation study, our proposed method seems to perform reasonably well.
%It can be readily seen that even with high percentage of zero-inflation, our proposed method seems to perform reasonably well.

\begin{table}
	\caption{Results of the simulation study with $15 \%$ zeros in both the response and covariate.}
	\centering
	\begin{tabular}{|c|c|c|c|c|c|c|c|}
		\hline
		\multicolumn{8}{|c|}{n=50}  \\
		\hline 
		Parameters& circular mean (c.d.) & Bias & CP & Parameters & mean (s.d.) & Relative Bias & CP\\
		\hline
		$\theta_{01}=0$& -0.019(0.193) & -0.019 & 0.71 & $r_1=0.9$& 0.907(0.083) & 0.008 & 0.90 \\ 
		$\theta_{02}=0$ & 0.019(0.128) & 0.019 & 0.74 & $r_2=1.2$ & 1.177(0.100)& -0.019 & 0.92 \\ 
		$\theta_{11}=0$& 0.020(0.208) & 0.020 & 0.70 & $\rho_1=0.85$& 0.845(0.030) & -0.006 & 0.95 \\
		$\theta_{12}=0$ & -0.018(0.110) & -0.018 & 0.75 & $\rho_2=0.85$ & 0.846(0.030) & -0.005 & 0.95 \\
		\hline \hline
		\multicolumn{8}{|c|}{n=100}  \\
		\hline 
		Parameters& circular mean (c.d.)& Bias & CP & Parameters & mean (s.d.) & Relative Bias & CP\\
		\hline
		$\theta_{01}=0$& -0.003(0.061) & -0.003 & 0.77 &
		$r_1=0.9$& 0.903(0.058) & 0.003  &0.91 \\
		$\theta_{02}=0$ & -0.004 (0.038) & -0.004 & 0.76
		& $r_2=1.2$ & 1.192(0.055) & 0.007 & 0.91 \\ 
		$\theta_{11}=0$& 0.002(0.066) & 0.002 & 0.77 &
		$\rho_1=0.85$& 0.847(0.022) &-0.004 & 0.92 \\
		$\theta_{12}=0$& 0.003(0.031) & 0.003 & 0.76 
		& $\rho_2=0.85$ & 0.848(0.020) & -0.002 & 0.95 \\
		\hline
		
	\end{tabular}
	
	\label{t1}
	
\end{table}

\begin{table}
	\caption{Results of the simulation study with $10 \%$ zeros in both the response and covariate.}
	\centering
	\begin{tabular}{|c|c|c|c|c|c|c|c|}
		\hline
		\multicolumn{8}{|c|}{n=50}  \\
		\hline 
		Parameters& circular mean (c.d.) & Bias & CP & Parameters & mean (s.d.) & Relative Bias & CP\\
		\hline
		$\theta_{01}=0$& 0.025(0.198) & 0.025 & 0.71 & $r_1=1.2$& 1.177(0.127) & -0.019 & 0.90 \\ 
		$\theta_{02}=0$ & -0.017(0.114) & -0.017 & 0.77 & $r_2=1.2$ & 1.183(0.087)& -0.014 & 0.91 \\ 
		$\theta_{11}=0$& -0.023(0.180) & -0.023 & 0.71 & $\rho_1=0.85$& 0.843(0.030) & -0.008 & 0.95 \\
		$\theta_{12}=0$ & 0.014(0.098) & 0.014 & 0.78 & $\rho_2=0.85$ & 0.844(0.030) & -0.007 & 0.95 \\
		\hline \hline
		\multicolumn{8}{|c|}{n=100}  \\
		\hline 
		Parameters& circular mean (c.d.)& Bias & CP & Parameters & mean (s.d.) & Relative Bias & CP\\
		\hline
		$\theta_{01}=0$& -0.005(0.044) & -0.005 & 0.79 & $r_1=1.2$& 1.189(0.080) & -0.009  &0.93 \\
		$\theta_{02}=0$ & -0.011 (0.046) & -0.011 & 0.76 & $r_2=1.2$ & 1.190(0.056) & -0.008 & 0.94 \\ 
		$\theta_{11}=0$& 0.005(0.039) & 0.005 & 0.78 & $\rho_1=0.85$& 0.847(0.021) & -0.004 & 0.95 \\
		$\theta_{12}=0$& 0.012(0.038) & 0.012 & 0.76 & $\rho_2=0.85$ & 0.847(0.020) & -0.004 & 0.95 \\
		\hline
		
	\end{tabular}
	
	\label{t2}
	
\end{table}

\begin{table}
	\caption{Results of the simulation study with $10 \%$ zeros in response only.}
	\centering
	\begin{tabular}{|c|c|c|c|c|c|c|c|}
		\hline
		\multicolumn{8}{|c|}{n=50}  \\
		\hline 
		Parameters& circular mean (c.d.) & Bias & CP & Parameters & mean (s.d.) & Relative Bias & CP\\
		\hline
		$\theta_{01}=0$& -0.025(0.127) & -0.025 & 0.79 & $r_1=0.9$& 0.888(0.078) & -0.013 & 0.91 \\ 
		$\theta_{02}=\frac{\pi}{2}$ & 1.552(0.087) & -0.019 & 0.77 & $r_2=1.5$ & 1.448(0.122)& -0.034 & 0.88 \\ 
		$\theta_{11}=0$& 0.025(0.138) & 0.025 & 0.80 & $\rho_1=0.85$& 0.843(0.030) & -0.008 & 0.95 \\
		$\theta_{12}=0$ & 0.016(0.062) & 0.016 & 0.79 & $\rho_2=0.85$ & 0.844(0.031) & -0.007 & 0.93 \\
		\hline \hline
		\multicolumn{8}{|c|}{n=100}  \\
		\hline 
		Parameters& circular mean (c.d.)& Bias & CP & Parameters & mean (s.d.) & Relative Bias & CP\\
		\hline
		$\theta_{01}=0$& -0.060(0.022) & -0.060 & 0.86 & $r_1=0.9$& 0.894(0.034) & -0.007  &0.93 \\
		$\theta_{02}=\frac{\pi}{2}$ & 1.587 (0.030) & 0.016 & 0.78 & $r_2=1.5$ & 1.481(0.082) & -0.013 & 0.91 \\ 
		$\theta_{11}=0$& 0.059(0.024) & 0.059 & 0.88 & $\rho_1=0.85$& 0.846(0.021) & -0.005 & 0.93 \\
		$\theta_{12}=0$& -0.013(0.021) & -0.013 & 0.77 & $\rho_2=0.85$ & 0.847(0.021) & -0.004 & 0.93 \\
		\hline
		
	\end{tabular}
	
	\label{t3}
	
\end{table}

\begin{table}
	\caption{Results of simulation study with $10 \%$ zeros in covariate only.}
	\centering
	\begin{tabular}{|c|c|c|c|c|c|c|c|}
		\hline
		\multicolumn{8}{|c|}{n=50}  \\
		\hline 
		Parameters& circular mean (c.d.) & Bias & CP & Parameters & mean (s.d.) & Relative Bias & CP\\
		\hline
		$\theta_{01}=\frac{\pi}{2}$&  1.543(0.075) & 0.028 & 0.72 & $r_1=0.9$& 0.897(0.086) & -0.003 & 0.92 \\ 
		$\theta_{02}=0$ & 0.009(0.095) & 0.009 & 0.81 & $r_2=1.5$ & 1.441(0.116)& -0.039 & 0.90 \\ 
		$\theta_{11}=0$& 0.029(0.165) & 0.029 & 0.75 & $\rho_1=0.85$& 0.844(0.031) & -0.007 & 0.95 \\
		$\theta_{12}=0$ & -0.009(0.070) & -0.009 & 0.81 & $\rho_2=0.85$ & 0.841(0.031) & -0.011 & 0.95 \\
		\hline \hline
		\multicolumn{8}{|c|}{n=100}  \\
		\hline 
		Parameters& circular mean (c.d.)& Bias & CP & Parameters & mean (s.d.) & Relative Bias & CP\\
		\hline
		$\theta_{01}=\frac{\pi}{2}$& 1.532(0.060) & -0.039 & 0.76 & $r_1=0.9$& 0.897(0.056) & -0.037  &0.91 \\
		$\theta_{02}=0$ & 0.008 (0.027) & 0.008 & 0.83 & $r_2=1.5$ & 1.484(0.083) & -0.011 & 0.91 \\ 
		$\theta_{11}=0$& 0.041(0.061) & 0.041 & 0.76 & $\rho_1=0.85$& 0.847(0.021) & -0.004 & 0.95 \\
		$\theta_{12}=0$& -0.008(0.019) & -0.008 & 0.83 & $\rho_2=0.85$ & 0.847(0.021) & -0.004 & 0.94 \\
		\hline
		
	\end{tabular}
	\label{t4}
	
\end{table}

\begin{table}
	\caption{Results of the simulation study without zero-inflation.}
	\centering
	\begin{tabular}{|c|c|c|c|c|c|c|c|}
		\hline
		\multicolumn{8}{|c|}{n=50}  \\
		\hline 
		Parameters& circular mean (c.d.) & Bias & CP & Parameters & mean (s.d.) & Relative Bias & CP\\
		\hline
		$\theta_{01}=\frac{\pi}{2}$&  1.559(0.012) & -0.025 & 0.95 & $r_1=0.3$& 0.303(0.049) & 0.010 & 0.94 \\ 
		$\theta_{02}=\frac{\pi}{2}$ & 1.572(0.001) & 0.009 & 0.94 & $r_2=0.3$ & 0.318(0.044)& 0.060 & 0.95 \\ 
		$\theta_{11}=0$& 0.030(0.013) & 0.030 & 0.96 & $\rho_1=0.85$& 0.842(0.031) & -0.009 & 0.95 \\
		$\theta_{12}=0$ & 0.007(0.037) & 0.007 & 0.94 & $\rho_2=0.85$ & 0.847(0.030) & -0.004 & 0.94 \\
		\hline \hline
		\multicolumn{8}{|c|}{n=100}  \\
		\hline 
		Parameters& circular mean (c.d.)& Bias & CP & Parameters & mean (s.d.) & Relative Bias & CP\\
		\hline
		$\theta_{01}=\frac{\pi}{2}$& 1.567(0.002) & -0.004 & 0.93 & $r_1=0.3$& 0.301(0.032) & 0.003  &0.92 \\
		$\theta_{02}=\frac{\pi}{2}$ & 1.568 (0.006) & -0.003 & 0.93 & $r_2=0.3$ & 0.308(0.029) & 0.027 & 0.93 \\ 
		$\theta_{11}=0$& 0.003(0.003) & 0.003 & 0.94 & $\rho_1=0.85$& 0.847(0.020) & -0.004 & 0.95 \\
		$\theta_{12}=0$& 0.004(0.016) & 0.004 & 0.91 & $\rho_2=0.85$ & 0.848(0.019) & -0.002 & 0.97 \\
		\hline
		
	\end{tabular}
	
	\label{t5}
	
\end{table}

\begin{table}
	\caption{Results of the simulation study with approximately $35 \%$ zeros in the response and $50 \%$ zeros in the covariate.}
	\centering
	\begin{tabular}{|c|c|c|c|c|c|c|c|}
		\hline
		\multicolumn{8}{|c|}{Model I} \\
		\hline 
		Parameters& circular mean (c.d.) & Bias & CP & Parameters & mean (s.d.) & Relative Bias & CP\\
		\hline
		$\theta_{01}=0.070$ &  0.073(0.004) & 0.003 & 0.56 & $r_1=0.9$& 0.898(0.060) & -0.002 & 0.91 \\ 
		$\theta_{02}=0.070$ & 0.063(0.001) & -0.007 & 0.47 & $r_2=1.2$ & 1.186(0.057)& -0.012 & 0.90 \\ 
		$\theta_{11}=0$ & -0.002(0.011) & -0.002 & 0.57 & $\rho_1=0.93$& 0.928(0.010) & -0.077 & 0.94 \\
		$\theta_{12}=0$ & -0.002(0.008) & -0.002 & 0.44 & $\rho_2=0.95$ & 0.949(0.008) & -0.001 & 0.95 \\
		\hline 
		\multicolumn{8}{|c|}{Model II}  \\
		\hline 
		Parameters& circular mean (c.d.)& Bias & CP & Parameters & mean (s.d.) & Relative Bias & CP\\
		\hline
		$\theta_{01}=0.070$ & 0.092(0.006) & 0.022 & 0.54 & $r_1=0.9$& 0.894(0.064) & -0.007  &0.88 \\
		$\theta_{02}=0.070$ & 0.036(0.001) & -0.034 & 0.09 & $r_2=1.2$ & 1.019(0.010) & -0.151 & 0.08 \\ 
		$\theta_{11}=0$& -0.046(0.007) & -0.046 & 0.55 & $\rho_1=0.93$& 0.922(0.011) & -0.084 & 0.92 \\
		$\theta_{12}=0$& -0.029(0.001) & -0.029 & 0.10 & $\rho_2=0.95$ & 0.990(0.003) & 0.042 & 0.17 \\
		\hline 
		
		\multicolumn{8}{|c|}{Model III}  \\
		\hline 
		Parameters& circular mean (c.d.)& Bias & CP & Parameters & mean (s.d.) & Relative Bias & CP\\
		\hline
		$\theta_{01}=0.070$ & 0.088(0.004) & 0.018 & 0.55 & $r_1=0.9$& 0.899(0.057) & -0.001  &0.89 \\
		$\theta_{02}=0.070$ & 0.037(0.001) & -0.033 & 0.10 & $r_2=1.2$ & 1.016(0.009) & -0.153 & 0.05 \\ 
		$\theta_{11}=0$& -0.020(0.005) & -0.020 & 0.55 & $\rho_1=0.93$& 0.928(0.011) & -0.077 & 0.96 \\
		$\theta_{12}=0$& -0.031(0.001) & -0.031 & 0.09 & $\rho_2=0.95$ & 0.991(0.003) & 0.043 & 0.14 \\
		\hline 
	\end{tabular}
	
	\label{t6}
	
\end{table}

\begin{table}
	\caption{Results of the simulation study under mis-specified model with $p=0.1$.}
	\centering
	\begin{tabular}{|c|c|c|c|c|c|c|c|}
		\hline
		\multicolumn{8}{|c|}{Model I} \\
		\hline 
		Parameters& circular mean (c.d.) & Bias & CP & Parameters & mean (s.d.) & Relative Bias & CP\\
		\hline
		$\theta_{01}=0.070$ &  0.075(0.003) & 0.005 & 0.51 & $r_1=0.9$& 0.901(0.064) & 0.001 & 0.90 \\ 
		$\theta_{02}=0.070$ & 0.060(0.001) & -0.010 & 0.48 & $r_2=1.2$ & 1.170(0.053)& -0.025 & 0.83 \\ 
		$\theta_{11}=0$ & -0.018(0.004) & -0.018 & 0.51 & $\rho_1=0.93$& 0.934(0.010) & 0.004 & 0.92 \\
		$\theta_{12}=0$ & -0.002(0.001) & -0.002 & 0.48 & $\rho_2=0.95$ & 0.955(0.007) & 0.005 & 0.84 \\
		\hline 
		\multicolumn{8}{|c|}{Model II}  \\
		\hline 
		Parameters& circular mean (c.d.)& Bias & CP & Parameters & mean (s.d.) & Relative Bias & CP\\
		\hline
		$\theta_{01}=0.070$ & 0.061(0.003) & -0.009 & 0.47 & $r_1=0.9$& 0.912(0.057) & 0.013  &0.79 \\
		$\theta_{02}=0.070$ & 0.035(0.001) & -0.035 & 0.02 & $r_2=1.2$ & 1.005(0.003) & -0.163 & 0.01 \\ 
		$\theta_{11}=0$& -0.033(0.004) & -0.033 & 0.46 & $\rho_1=0.93$& 0.937(0.010) & 0.008 & 0.82 \\
		$\theta_{12}=0$& -0.033(0.001) & -0.033 & 0.02 & $\rho_2=0.95$ & 0.998(0.001) & 0.051 & 0.02 \\
		\hline 
		
		\multicolumn{8}{|c|}{Model III}  \\
		\hline 
		Parameters& circular mean (c.d.)& Bias & CP & Parameters & mean (s.d.) & Relative Bias & CP\\
		\hline
		$\theta_{01}=0.070$ & 0.056(0.003) & -0.014 & 0.46 & $r_1=0.9$& 0.911(0.057) & 0.011  &0.87 \\
		$\theta_{02}=0.070$ & 0.035(0.001) & -0.035 & 0.02 & $r_2=1.2$ & 1.005(0.009) & -0.163 & 0.00 \\ 
		$\theta_{11}=0$& -0.003(0.004) & -0.003 & 0.48 & $\rho_1=0.93$& 0.935(0.011) & 0.005 & 0.90 \\
		$\theta_{12}=0$& -0.033(0.001) & -0.033 & 0.02 & $\rho_2=0.95$ & 0.998(0.003) & 0.051 & 0.03 \\
		\hline 
	\end{tabular}
	
	\label{t7}
	
\end{table}

\begin{table}
	\caption{Results of the simulation study under mis-specified model with $p=0.2$.}
	\centering
	\begin{tabular}{|c|c|c|c|c|c|c|c|}
		\hline
		\multicolumn{8}{|c|}{Model I} \\
		\hline 
		Parameters& circular mean (c.d.) & Bias & CP & Parameters & mean (s.d.) & Relative Bias & CP\\
		\hline
		$\theta_{01}=0.070$ &  0.065(0.002) & -0.005 & 0.51 & $r_1=0.9$& 0.919(0.067) & 0.021 & 0.87 \\ 
		$\theta_{02}=0.070$ & 0.056(0.001) & -0.014 & 0.51 & $r_2=1.2$ & 1.141(0.051)& -0.049 & 0.72 \\ 
		$\theta_{11}=0$ & -0.016(0.003) & -0.016 & 0.51 & $\rho_1=0.93$& 0.940(0.009) & 0.011 & 0.73 \\
		$\theta_{12}=0$ & -0.007(0.001) & -0.007 & 0.51 & $\rho_2=0.95$ & 0.961(0.007) & 0.012 & 0.61 \\
		\hline 
		\multicolumn{8}{|c|}{Model II}  \\
		\hline 
		Parameters& circular mean (c.d.)& Bias & CP & Parameters & mean (s.d.) & Relative Bias & CP\\
		\hline
		$\theta_{01}=0.070$ & 0.055(0.001) & -0.015 & 0.33 & $r_1=0.9$& 0.950(0.040) & 0.056  & 0.54 \\
		$\theta_{02}=0.070$ & 0.034(0.001) & -0.036 & 0.00 & $r_2=1.2$ & 1.002(0.001) & -0.165 & 0.00 \\ 
		$\theta_{11}=0$& -0.044(0.001) & -0.044 & 0.34 & $\rho_1=0.93$& 0.963(0.008) & 0.035 & 0.39 \\
		$\theta_{12}=0$& -0.033(0.001) & -0.033 & 0.00 & $\rho_2=0.95$ & 0.999(0.001) & 0.052 & 0.00 \\
		\hline 
		
		\multicolumn{8}{|c|}{Model III}  \\
		\hline 
		Parameters& circular mean (c.d.)& Bias & CP & Parameters & mean (s.d.) & Relative Bias & CP\\
		\hline
		$\theta_{01}=0.070$ & 0.061(0.002) & -0.009 & 0.52 & $r_1=0.9$& 0.924(0.063) & 0.027  &0.84 \\
		$\theta_{02}=0.070$ & 0.033(0.001) & -0.037 & 0.00 & $r_2=1.2$ & 1.002(0.001) & -0.165 & 0.00 \\ 
		$\theta_{11}=0$& -0.011(0.003) & -0.011 & 0.50 & $\rho_1=0.93$& 0.941(0.009) & 0.012 & 0.70 \\
		$\theta_{12}=0$& -0.032(0.001) & -0.032 & 0.00 & $\rho_2=0.95$ & 0.999(0.001) & 0.052 & 0.00 \\
		\hline 
	\end{tabular}
	
	\label{t8}
	
\end{table}

\subsection{Model Comparison and Sensitivity Analysis}

For the purpose of comparison, we consider three different models, Model I, Model II and Model III, given by (\ref{reg1}) and (\ref{reg2}), (\ref{reg12}) and (\ref{reg22}), and (\ref{reg1}) and (\ref{reg22}), respectively. Note that Model I accounts for zero-inflation in both the response and the covariate, however, Model II does not account for zero-inflation, and Model III accounts for zero-inflation in the response only.
%We compare the performance of the proposed model (Model I), given by (\ref{reg1}) and (\ref{reg2}), with the two-stage circular-circular regression model, which does not account for zero-inflation (Model II), given by (\ref{reg12}) and (\ref{reg22}), and another model which accounts for zero-inflation in the response only (Model III), given by (\ref{reg1}) and (\ref{reg22}).
%\textcolor{red}{As mentioned before, when $\delta_X=\delta_Y=0$, the proposed model reduces to conventional two-stage circular-circular regression. One can also consider zero inflation only in the response variable just by considering $\delta_X=0$. 
We first compare the performance of parameter estimates associated with Model I, Model II and Model III when the data is generated from Model I with sample size $n=100$ and $\delta_X=\delta_Y=0.070$ radians ($\ang{4}$). We consider the priors as discussed in the Subsection \ref{SA} with $a_{\rho_i}=2$ for $i=1,2$. The averages of the estimates over 500 replications are reported in Table \ref{t6}. It is observed that the performance of most of the estimates based on Model II and Model III are biased compared to those of Model I. The estimates of the parameters associated with the regression of $X$ on $V$ based on Model II and Model III have large bias and very low CP. It is important to note that the estimates associated with the regression of $Y$ on $X$ based on Model III, which incorporates zero-inflation in the response, are biased compared to those of Model I that also incorporates zero-inflation in the covariate. These observations clearly indicate that both Model II and Model III are inadequate to model the underlying relationship in the data when there is significant zero-inflation in both response and covariate.

In order to carry out a sensitivity analysis, we consider a misspecified simulation model, where $\theta_Y$ and $\theta_X$ are first generated as before and then contaminated with randomly occurring zeros with probability $p$. Note that the aforementioned model reduces to the proposed two-stage circular-circular regression model, given by (\ref{reg1})-(\ref{reg2}), for $p=0$. We generate data of sample size $n=100$  for two different choices of $p$ with $\delta=0.070$, and compare the different models with respect to bias and CP based on 500 replications. The results are presented in Table \ref{t7} and \ref{t8}. It is observed that the performance of the estimates for all the models deteriorate with respect to both bias and CP as $p$ increases. The bias and CP for most of the parameters based on Model I are considerably better compared to Model II and III. In particular, the CP of the parameters associated with the  regression of $X$ on $V$ based on Model II and Model III are close to zero. Even with the higher value of $p$, the estimates obtained from Model I are robust compared to those of Model II and Model III. 

Next, we compare the run-time of different models per dataset in Table \ref{Timetable}, under the simulation settings presented in Tables \ref{t6}-\ref{t8}, using a computer
equipped with 32 GB RAM and 2.59 GHz Intel Core(TM) i7-8850H processor.
As expected, the computing time increases with the increase in the proportion of zeros for Model I and Model III. We also observed the same pattern for the simulation settings corresponding to Tables \ref{t1}-\ref{t5}. The difference among the run-times for the three different models is less than 3 minutes, even with large proportion of zeros. As it turns out, Model I seems to have an edge concerning the trade-off between computational time and model performance in presence of zero-inflation.

\begin{table}
	\caption{Computation Time in minutes.}
	\centering
	\begin{tabular}{|c|c|c|c|c|}
		\hline
		
		& Model I & Model II & Model III \\
		\hline
		Table 6 & 6.75 & 4.95 & 5.01 \\
		Table 7 & 6.95 & 4.82 & 5.81 \\
		Table 8 & 7.66 & 5.21 & 5.95 \\
		\hline 
	\end{tabular}\label{Timetable}

\end{table}

\section{Case Studies}\label{Case}	
As discussed before, applications of circular-circular regression are commonly found in the domain of orthopaedics and ophthalmology. In this Section, we consider two different case studies from each of these application areas. Significant amount of zero-inflation is observed in both `Cataract Surgery Data' and `Abnormal Gait Data'.
%To compare the results obtained from our proposed model (Model I), we also apply the two-stage circular-circular regression without zero-inflation (Model II), and zero-inflation in the response only (Model III), for both data analysis. 
The results from analysis of these datasets based on the proposed methodology along with the practical implications are discussed below.

\subsection{Analysis of Cataract Surgery Data}\label{Data}
As discussed in Section \ref{Intro}, we consider a dataset on astigmatism observed at three different inspections from a study on cataract surgery conducted at Disha Eye Hospital and Research Center.  
%\textcolor{green}{As discussed before, patients are inspected at regular intervals for monitoring visual recovery and minimizing surgical trauma. Therefore, it is of interest to model the axis of astigmatism after two weeks of the surgery for the purpose of identifying patients who require additional care. \citet{Jhatr1705} fitted a circular-circular regression model considering the response and the covariate as the axis of astigmatism after 15 days and 7 days of the surgery, respectively, where only the part of the dataset was considered in which the responses were zero-inflated but the covariates were not. In this analysis, we consider the dataset corresponding to \textcolor{green}{Snare, Vectis and Conventional Phacoemulsification techniques of cataract surgery, where both the response and the covariate are zero-inflated.}
In the proposed two-stage setup, we first consider the response ($\theta_Y$) and the covariate ($\theta_X$) as the axis of astigmatism after 15 days and 7 days of the surgery, respectively. Among the 54 observations, there are $31\%$ and $35\%$ zeros in response and covariate, respectively. In this study, measurements on the axis of astigmatism just after a day of the surgery are also available. Next, we model the axis of astigmatism after 7 days of the surgery ($\theta_X$) with covariate as the axis of astigmatism just after a day of surgery ($\theta_{V}$). In order to apply the methodology provided in  Section \ref{method}, we consider $\delta_X=\delta_Y=0.035$ radians ($\ang{2}$) and $a_{\rho_i}=2$ for $i=1,2$. This particular choice of $\delta_X$ and $\delta_Y$ is considered  as the original axes of astigmatism are censored in the interval $(-\ang{2},\ang{2})$.

We generate 250,000 samples from the posterior distributions of the associated model parameters using the MCMC algorithm and find the posterior mean/circular mean and s.d./c.d. based on every 10th iterate discarding the first 70,000 iterations as burn-in. The convergence of the chains is monitored graphically and using Geweke's diagnostic test after taking cosine transformation of the samples for the circular parameters. The mean/circular mean, s.d/c.d. of the parameters and the results for Geweke's diagnostic test are reported in Appendix A of the Supplementary Material. We also report the $95 \%$ highest posterior density (HPD) credible interval. The algorithm for finding the HPD credible interval for circular parameters is provided in \citet{jhatr2017rg}. As expected, the $95 \%$ HPD credible interval for both $r_1$ and $r_2$ does not contain 1. This indicates that $\theta_Y$ and $\theta_{X}$ are dependent on $\theta_{X}$ and $\theta_{V}$, respectively.
%It is evident that Model I and its special case fit the data better compared to both Model II and Model III with respect to AIC and BIC (See Appendix A in the Supplementary Material). 
%	To asses the adequacy of M\"obius transformation based models, we present a modified probability-probability (MPP) plot for all the models in Figure \ref{PP}. In MPP plot, we consider Kaplan-Maier estimate instead of the empirical probabilities to incorporate the censored residuals. Note that, we have taken 0 as the origin and represent all the residuals in the interval $[0,2\pi)$ for the computation of Kaplan-Maier estimate. It is visible from the MPP plots that the proposed models (Model I and its special case),  which incorporate zero-inflation in the covariate, provide better fit compared to Model II and Model III. 
%Henceforth, we only consider Model I and its special case for further analysis.

\begin{figure}[htp]
	\centering
	\subfigure [\scriptsize Astigmatism at day 7 based on day 1] {
		\includegraphics[width=5.5cm,height=5.5cm]{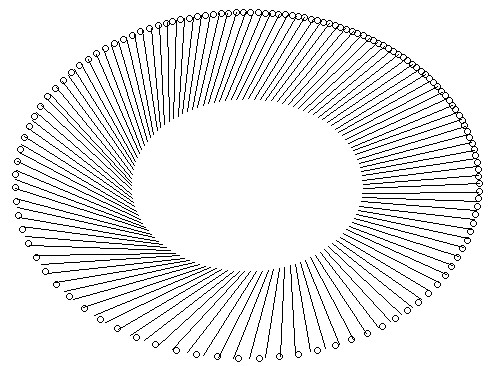} 
		\label{spoplo:subfig1}
	}
	\subfigure[\scriptsize Astigmatism at day 15 based on day 7]{
		\includegraphics[width=5.5cm,height=5.5cm]{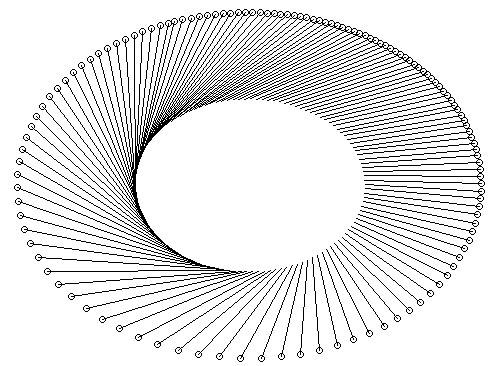} 
		\label{spoplo:subfig2}
	}
	\caption{Comparison of the recovery processes in the first and the second week after surgery based on Model I. }
	\label{spom1}
\end{figure}

\begin{figure}[htp]
	\centering
	\subfigure [\scriptsize Astigmatism at day 7 based on day 1] {
		\includegraphics[width=5.5cm,height=5.5cm]{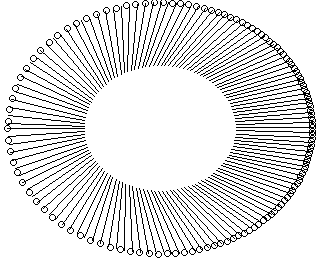} 
		\label{spoplo2:subfig1}
	}
	\subfigure[\scriptsize Astigmatism at day 15 based on day 7]{
		\includegraphics[width=5.5cm,height=5.5cm]{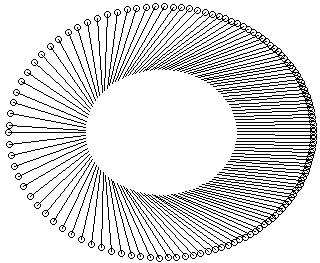} 
		\label{spoplo2:subfig2}
	}
	\caption{Comparison of the recovery processes in the first and the second week after surgery based on the special case of Model I.}
	\label{sporm}
\end{figure}

In order to compare the recovery processes in consecutive weeks, we present spoke plots in Figure \ref{spom1} and Figure \ref{sporm} based on the estimates from Model I and its special case, respectively. In the spoke plots, the predicted axes of astigmatism in the outer circle are joined with the corresponding axes of astigmatism a week before in the inner circle. It is important to note that there is a preferred direction for the axis of astigmatism, and it makes practically more sense to consider the equivalence of symmetric points about $\ang{0}$. It can be seen from the spoke plots based on Model I (See Figure \ref{spom1}), that the predicted responses for symmetric observations around $\ang{0}$ are not symmetric. For example, the predicted mean direction for $\ang{90}$ is rotated towards $\ang{0}$, while for $\ang{270}$, it is rotated away from $\ang{0}$.  However, for the special case, the symmetric points rotate symmetrically towards the preferred direction (See Figure \ref{sporm}). Hence, the equivalence with respect to the preferred direction is maintained. Also, the predicted axes of astigmatism in the second week (astigmatism at day 15 based on day 7) are more attracted towards $\ang{0}$ than those of the first week (astigmatism at day 7 based on day 1) for both the models. This indicates that the recovery in the second week is more than the recovery in the first week. This finding is also supported by the fact that $Pr[\beta_1-b_1 > 0 | D]=0.96$, based on the estimates obtained from the special case of Model I.

\begin{figure}
	\centering
	
	\begin{tabular}{cccc}
		\text{\scriptsize Astigmatism at day 7 on day 1} & \text{\scriptsize Astigmatism at day 15 on day 1} & \text{\scriptsize Astigmatism at day 15 on day 7} & \\
		
		\includegraphics[width=4cm,height=4cm]{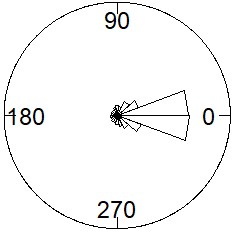} &
		\includegraphics[width=4cm,height=4cm]{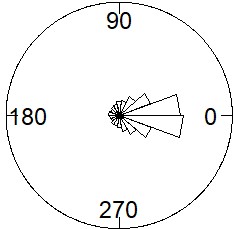} & 
		\includegraphics[width=4cm,height=4cm]{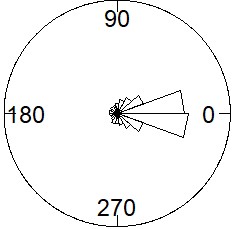} & \text{\scriptsize Initial axis $\ang{0}$} \\
		\includegraphics[width=4cm,height=4cm]{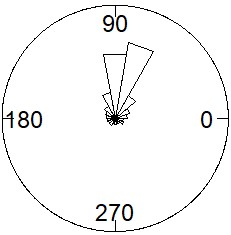} &
		\includegraphics[width=4cm,height=4cm]{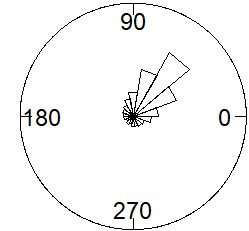} &
		\includegraphics[width=4cm,height=4cm]{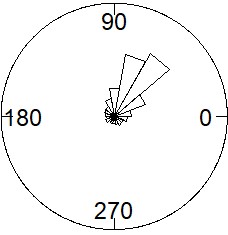} & \text{\scriptsize Initial axis $\ang{90}$} \\
		\includegraphics[width=4cm,height=4cm]{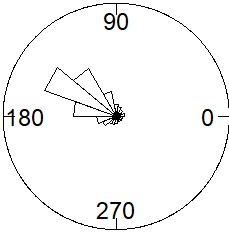} &
		\includegraphics[width=4cm,height=4cm]{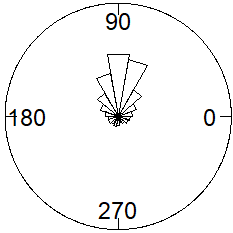} &
		\includegraphics[width=4cm,height=4cm]{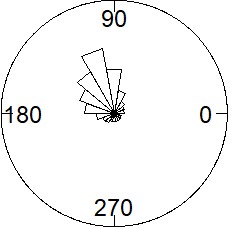} & \text{\scriptsize Initial axis $\ang{180}$} \\
		
	\end{tabular}
	\caption{Rose diagrams for posterior predictive distribution  based for three different initil conditions: normal ($\ang{0}$), intermediate ($\ang{90}$), and serious case ($\ang{180}$) of astigmatism.}
	\label{PPPR}
\end{figure}

Medical practitioners are more interested in identifying patients whose performances either improve or deteriorate. For this purpose, we consider Model I and compute the posterior predictive distributions of the axis of astigmatism at day 15 and day 7 with three different initial conditions fixed at $\ang{0}$ (normal), $\ang{90}$ (intermediate), and $\ang{180}$ (serious) on day 1. We also compute the posterior predictive distribution of the axis of astigmatism at day 15 based on those three initial conditions on day 7. In Figure \ref{PPPR}, we present rose diagrams based on the aforementioned posterior predictive distributions to analyse the recovery of patients over time. In all the cases, it is seen that the predictive distribution is concentrated around $\ang{0}$ if the initial axis of astigmatism is close to $\ang{0}$ (See the top panel of Figure \ref{PPPR}). For both intermediate and serious cases, improvement is more prominent in the second week compared to the first week as the posterior predictive distribution shifts more towards $\ang{0}$ in the second week (See middle and bottom panel of Figure \ref{PPPR}). For the serious case, the patients who show early sign of recovery improve more at the end of two weeks than those with delayed recovery (See second and third column at the bottom panel in Figure \ref{PPPR}).

To compare the predicted values with the observed values, we convert the predicted values in degrees and round off to the nearest integer divisible by 4. In general, the patients not affected by astigmatism after 7 days of the surgery remain unaffected in the near future. In this dataset, 17 patients remain unaffected by astigmatism during the study period and the fitted model identifies all of them correctly. We detect the improvement in 17 out of the 21 patients. However, the deterioration is detected for only 6 out of the 13 patients. Therefore, one can conclude that the patients whose conditions have improved and seem to stay good in the future require less monitoring while the rest of the patients need more frequent monitoring and care.

%	\begin{figure}
%		\centering

%		\includegraphics[scale=0.35]{PP_Model1.pdf}
%		\includegraphics[scale=0.35]{PP_Model2.pdf}
%		\includegraphics[scale=0.35]{PP_Model3.pdf}
%       \includegraphics[scale=0.34]{Reduced_MPP2.pdf}	
% \caption{MPP plot for Model I (top left), 
%		Model II (top right), Model III (bottom left)          and Special case of Model I (bottom right).}
%		\label{PP}
%	\end{figure}

%\includegraphics[scale=0.35]{Reduced_MPP2.pdf}

%\usepackage[font={small,it}]{caption}

%\multicolumn{3}{|c|}{\quad \quad\quad \quad\quad \quad\quad Special Case} \\
%\hline
%Parameters & mean/circular mean (s.d./c.d.) & $95 \%$ HPD Credible Interval \\
%\hline
%	$r_1$ & 0.287 (0.094) & (0.110,0.433) \\ 
%$r_2$ & 0.123 (0.151) & (0.047,0.197) \\
%	$r_1-r_2$ & 0.164 (0.176) & (-0.030,0.348) \\
%$\rho_1$ & 0.843 (0.039) & (0.760,0.906) \\ 
%$\rho_2$ & 0.883 (0.047) & (0.807,0.933) \\ 
%\hline 
%BIC  & \multicolumn{2}{c|}{373.674} \\

\subsection{Analysis of Abnormal Gait Data}\label{Gait_Data}
We consider a study on gait control conducted at Bio-Engineering Unit, University of Calcutta. The objective of this study is to analyse the angular movements of the patient’s affected limb with that of a healthy person while keeping the angular movement of the healthy limb fixed at a specified level within the range of natural variation. For this purpose, we consider the response ($\theta_Y$) and the covariate ($\theta_X$) as the angle of ankle extension of the affected and the unaffected legs, respectively, of the 19 individuals with unilateral orthopaedic impairment. We also jointly model the angle of ankle extension of the non-dominated leg ($\theta_Y$) as the response with that of the dominated leg ($\theta_X$) as covariate based on the measurements taken from 10 healthy individuals. As mentioned before, more than $50\%$ of the observations are recorded as $\ang{0}$ in both response and covariate. For the purpose of joint modeling, we consider a shared-parameter model replacing $\beta_0$ with $\beta_0\gamma^{W}$ in (\ref{reg1}), where $\gamma \in \{z:z \in \mathbbm{C};|z|=1\}$, and $W$ takes value $1$ for orthopaedically impaired individuals, and $0$, otherwise. In order to apply the methodology provided in Section \ref{method}, we consider $|b_1|=1$, and $\delta_X=\delta_Y=0.0087$ radians ($\ang{0.5}$).
% \textcolor{blue}{Note that there is natural variation in the angles of ankle extensions between both the legs for a healthy individuals, but the difference is not significant \citep{Ankle_normal}. To incorporate this information, we consider the prior density of the parameter $\theta_{01}$ as von Mises with mean direction 0 and concentration parameter 800, and the prior for $r_1$ as $\pi(r_1)\propto e^{-r_1^2}$.}
We generate 150,000 samples from the posterior distributions of the associated model parameters using the MCMC algorithm and discard the first 50,000 iterations as burn-in. Here also, no convergence issues are found based on Geweke's diagnostic test. The parameter estimates along with $95 \%$ HPD credible interval are reported in Appendix A of the Supplementary Material. 
%The AIC and BIC values for Model I are smaller compared to those of the Model II and Model III.

\begin{figure}
	\centering
	\includegraphics[scale=0.6]{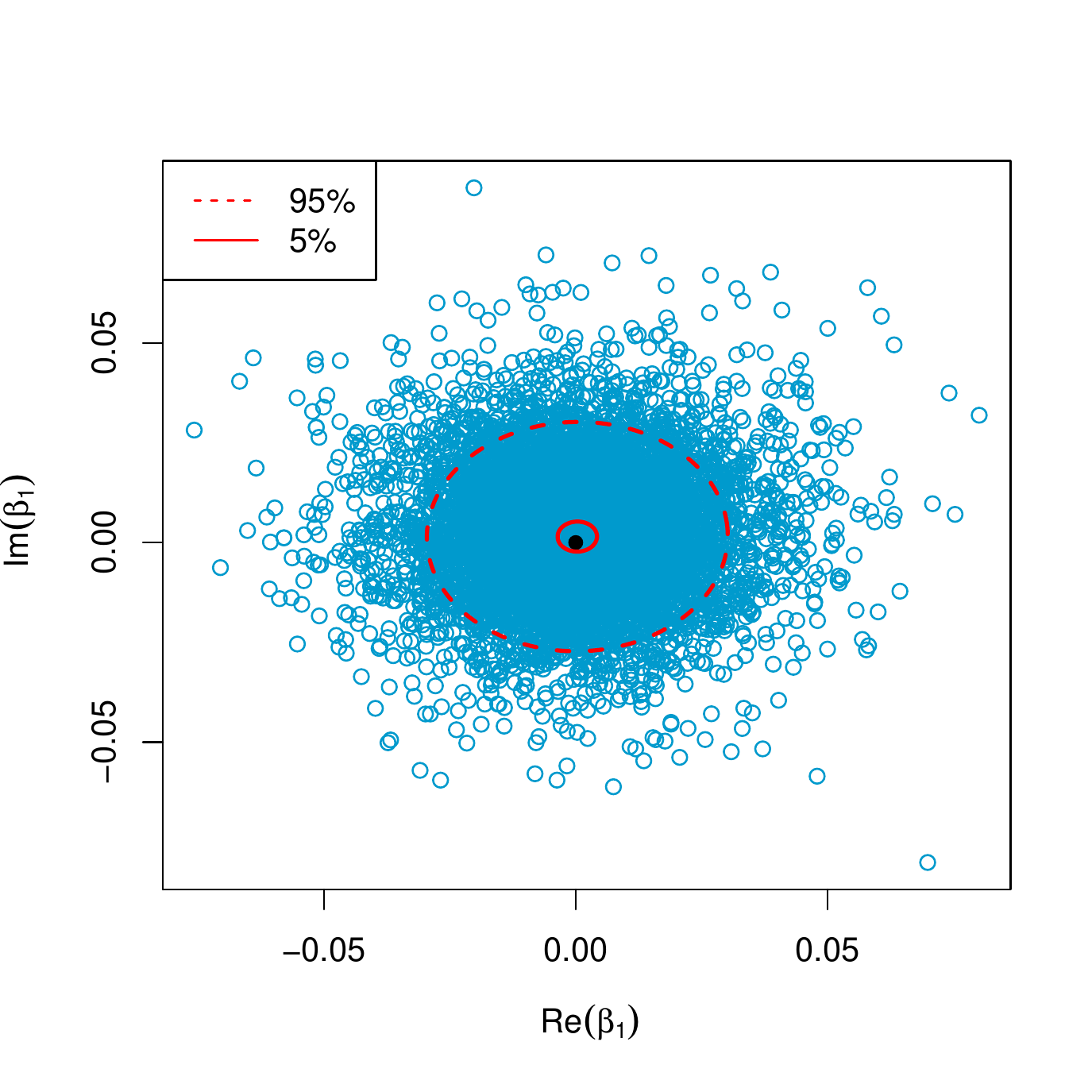}
	\caption{Confidence ellipses at $5\%$ and $95\%$ levels of the joint posterior distribution of $Re(\beta_1)$ and $Im(\beta_1)$ containing $(0,0)$.}
	\label{cont_beta1}
\end{figure}

\begin{figure}[htp]
	\centering
	\includegraphics[scale=0.6]{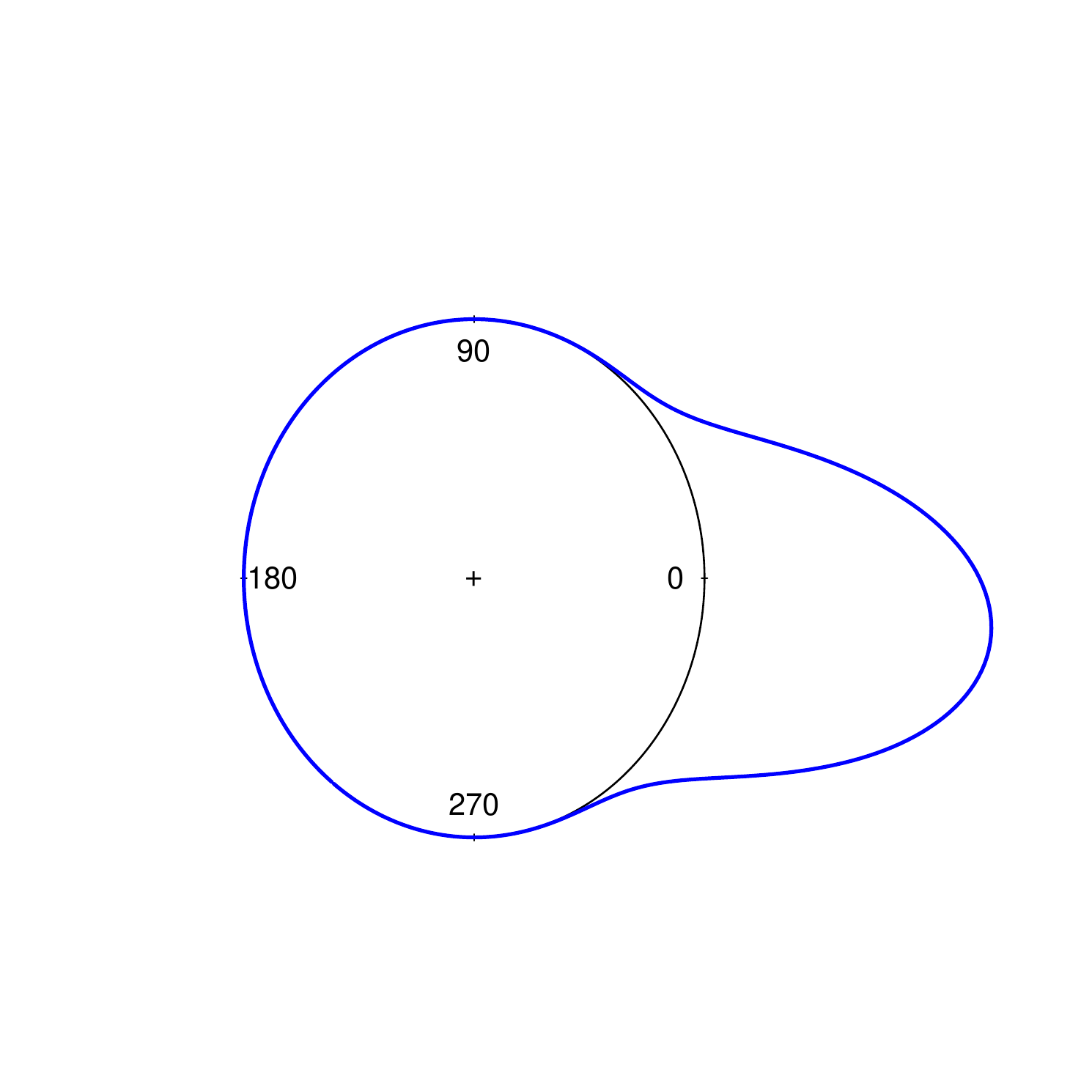} 
	\caption{Posterior density of $\theta_{\gamma}$.}
	\label{den_gam}
\end{figure}

\begin{figure}[htp]
	\centering
	\includegraphics[scale=0.5]{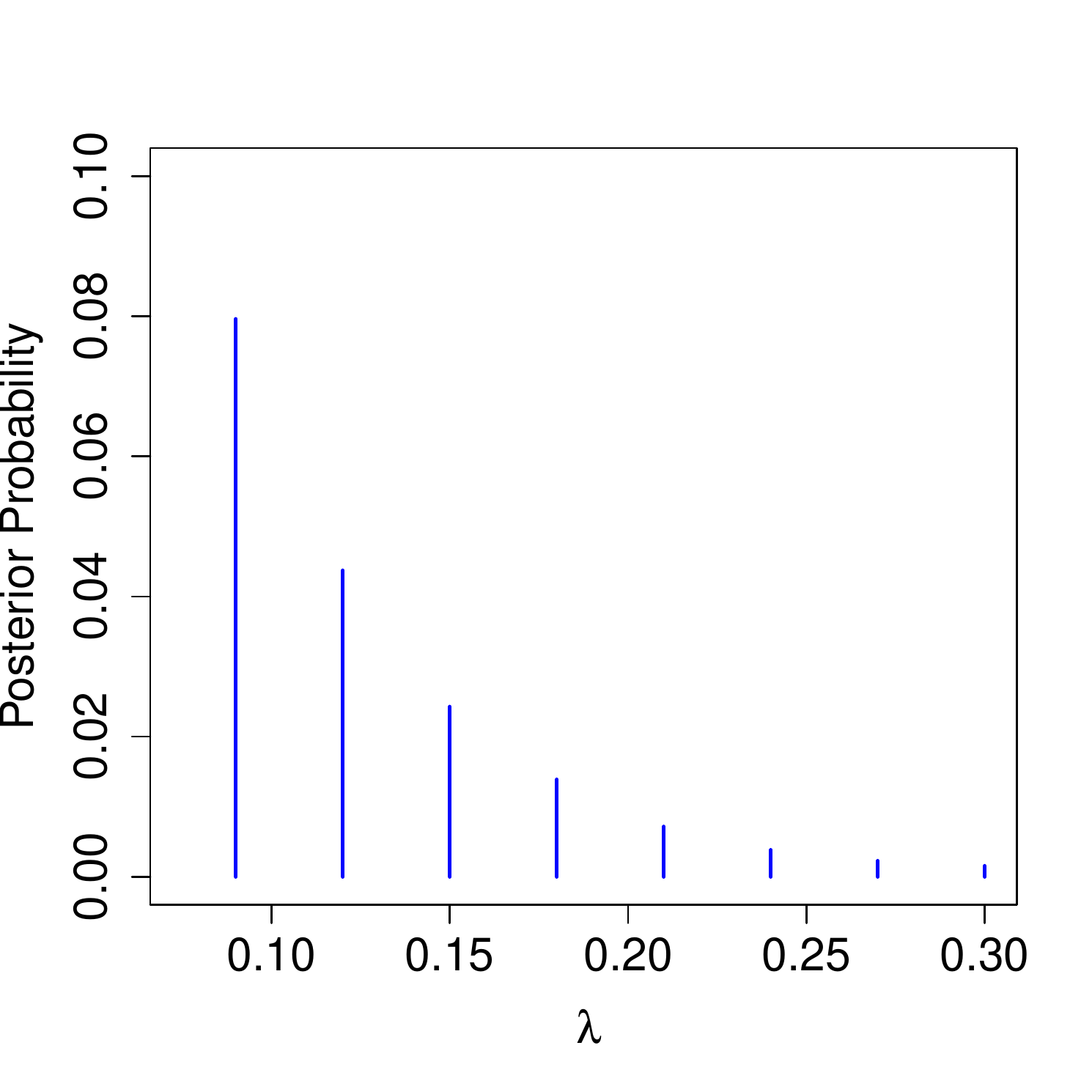} 
	\caption{Posterior probability $P\left[ \cos(\theta_{\gamma})< 1-\lambda|D \right]$ for different
		choices of $\lambda$.}
	\label{gam_test}
\end{figure}

\begin{figure}[htb]
	\centering
	\subfigure [\tiny Angle of ankle extension of the unaffected/dominating leg is $\ang{0}$] {
		\includegraphics[scale=0.45]{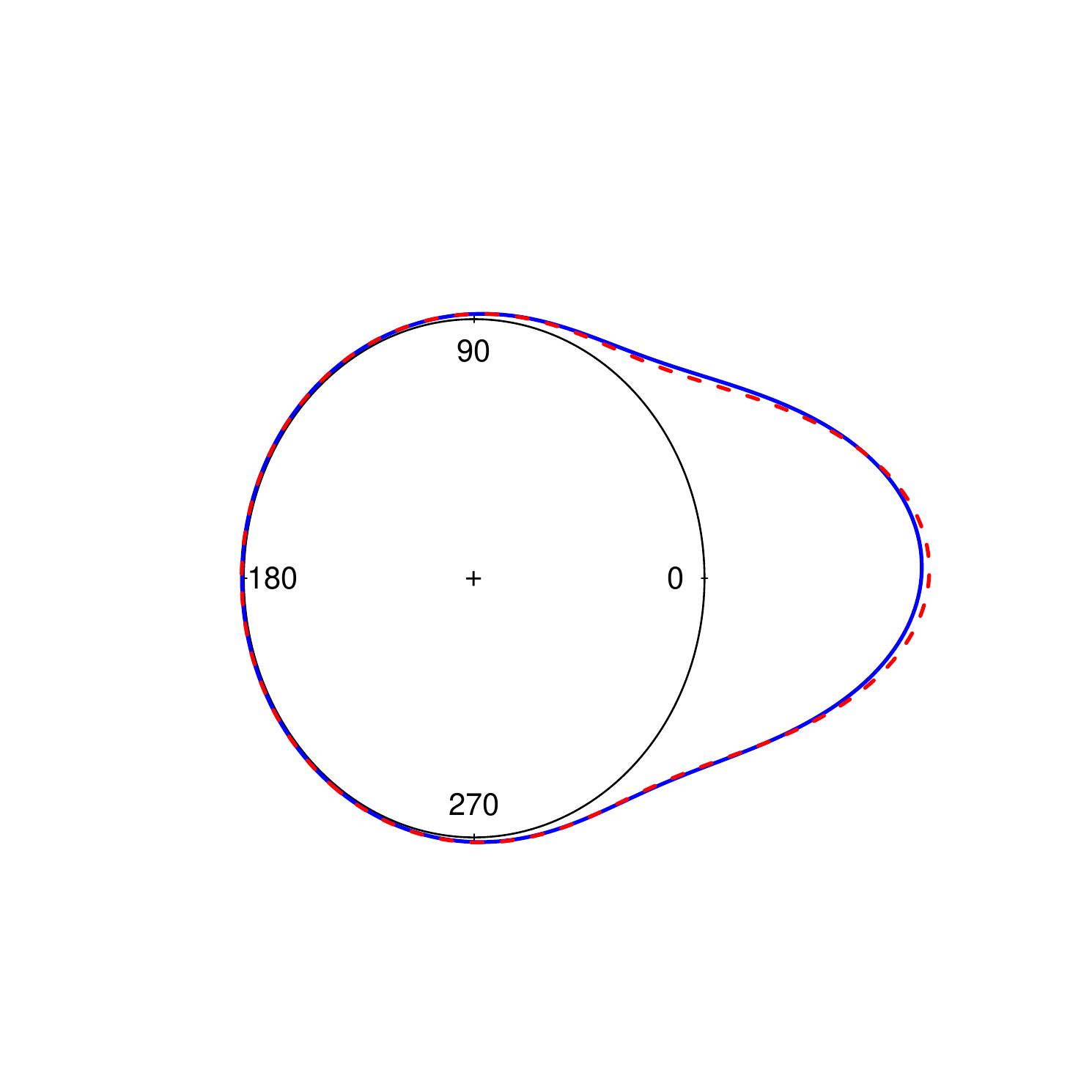} 
		\label{Ankle_den:subfig1}
	}
	\subfigure[\tiny Angle of ankle extension of the unaffected/dominating leg is $\ang{5}$]{
		\includegraphics[scale=0.45]{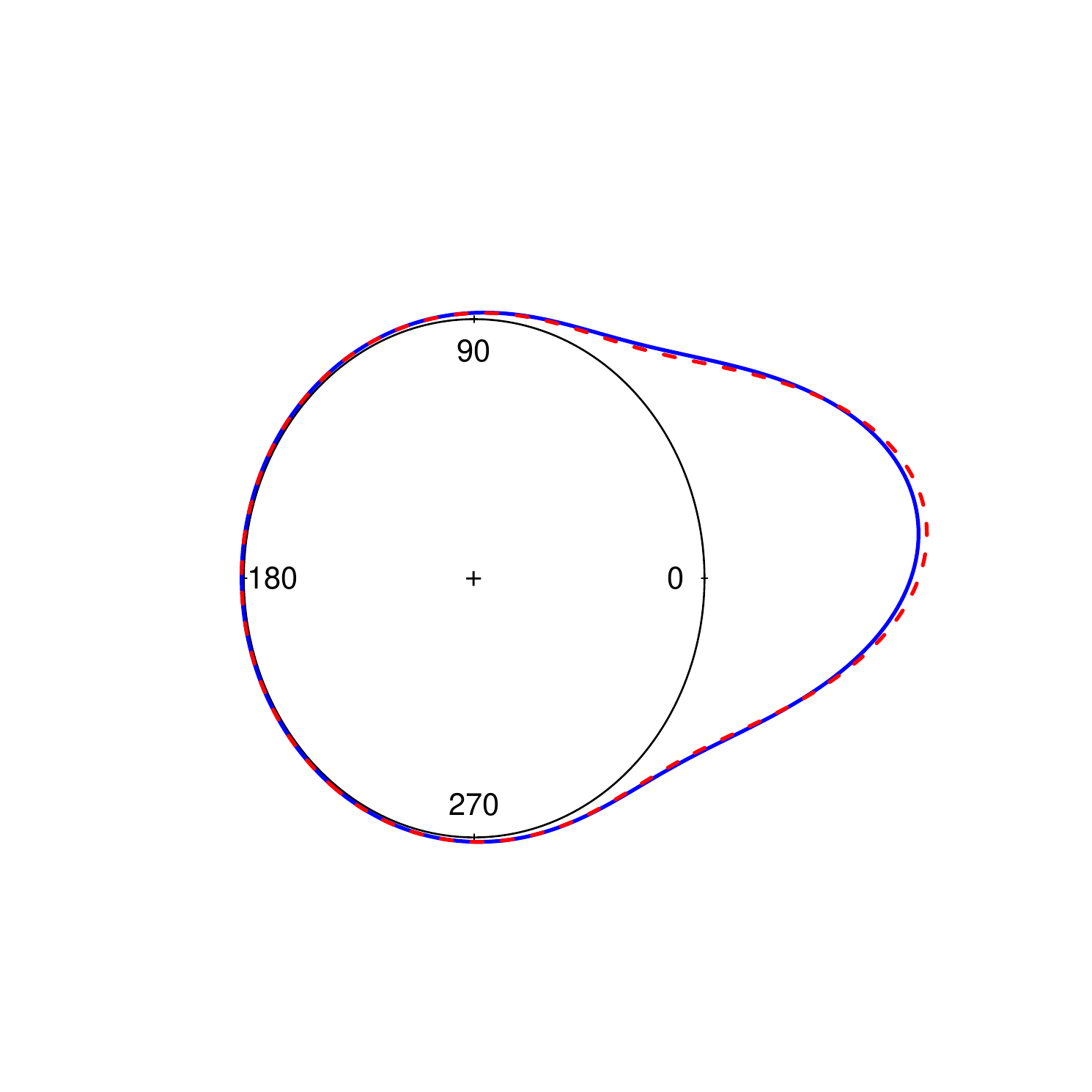} 
		\label{Ankle_den:subfig2}
	}
	\subfigure[\tiny Angle of ankle extension of the unaffected/dominating leg is $\ang{10}$]{
		\includegraphics[scale=0.45]{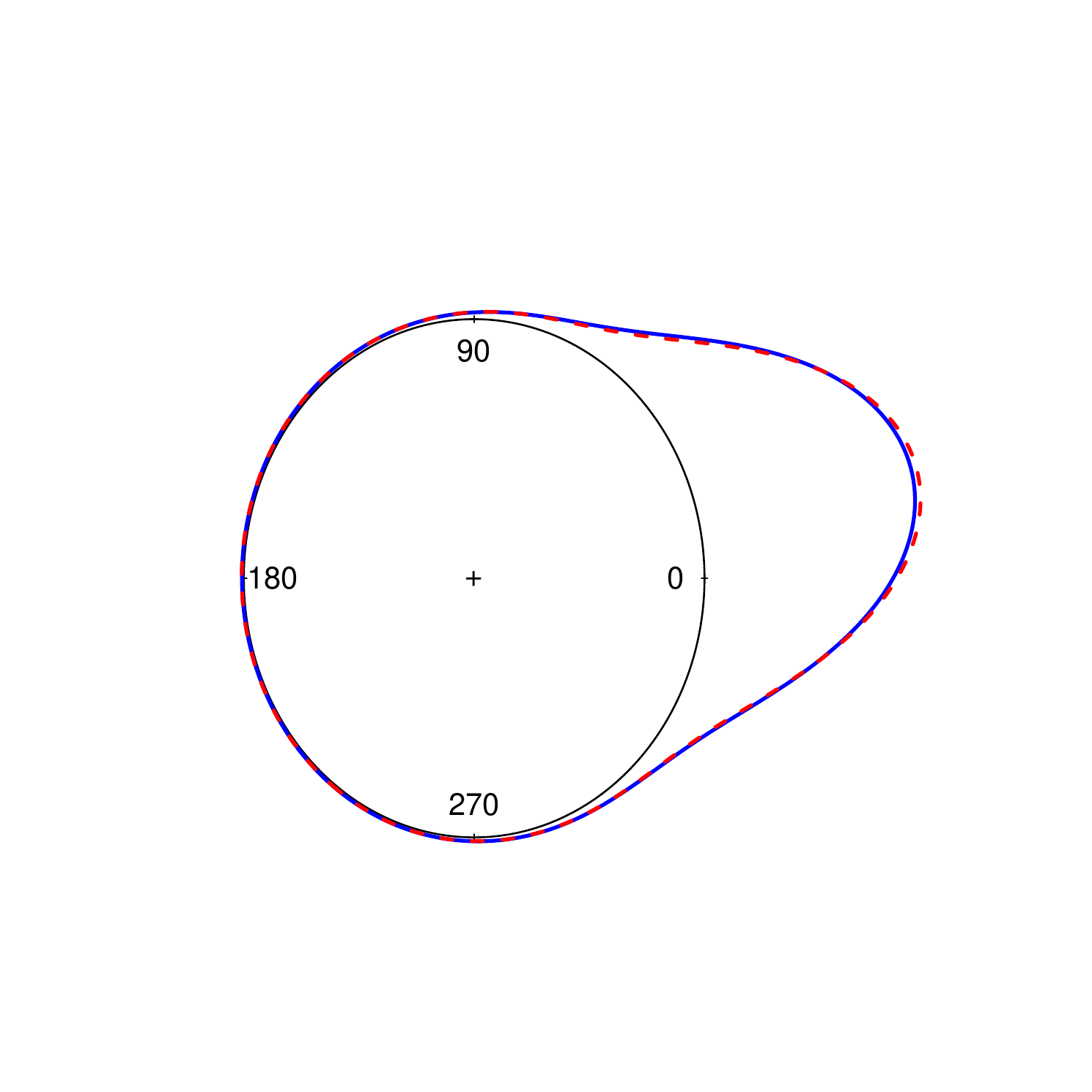} 
		\label{Ankle_den:subfig3}
	}
	\subfigure[\tiny Angle of ankle extension of the unaffected/dominating leg is $\ang{15}$]{
		\includegraphics[scale=0.45]{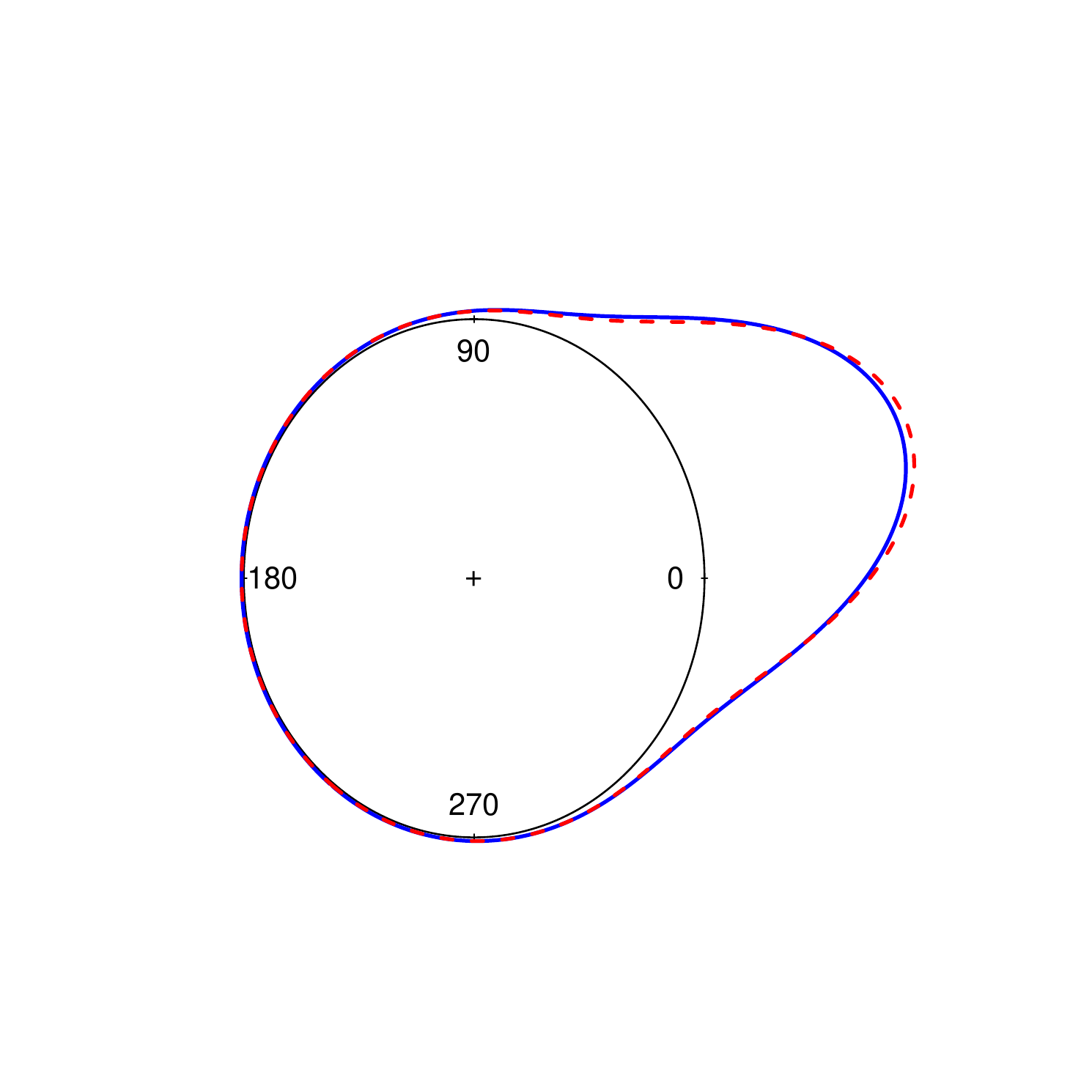} 
		\label{Ankle_den:subfig4}
	}
	\subfigure[\tiny Angle of ankle extension of the unaffected/dominating leg is $\ang{20}$]{
		\includegraphics[scale=0.45]{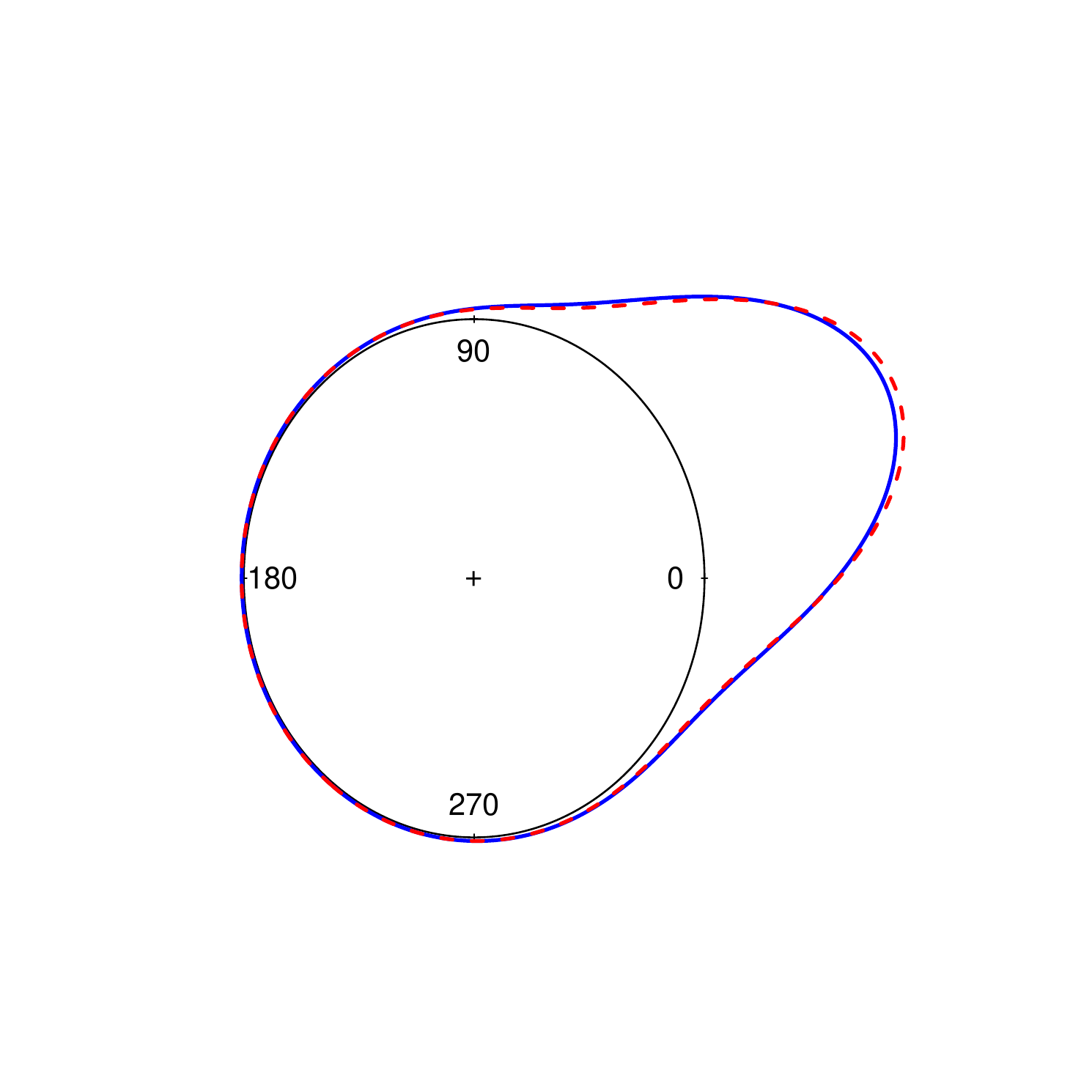} 
		\label{Ankle_den:subfig5}
	}
	\subfigure[\tiny Angle of ankle extension of the unaffected/dominating leg is $\ang{25}$]{
		\includegraphics[scale=0.45]{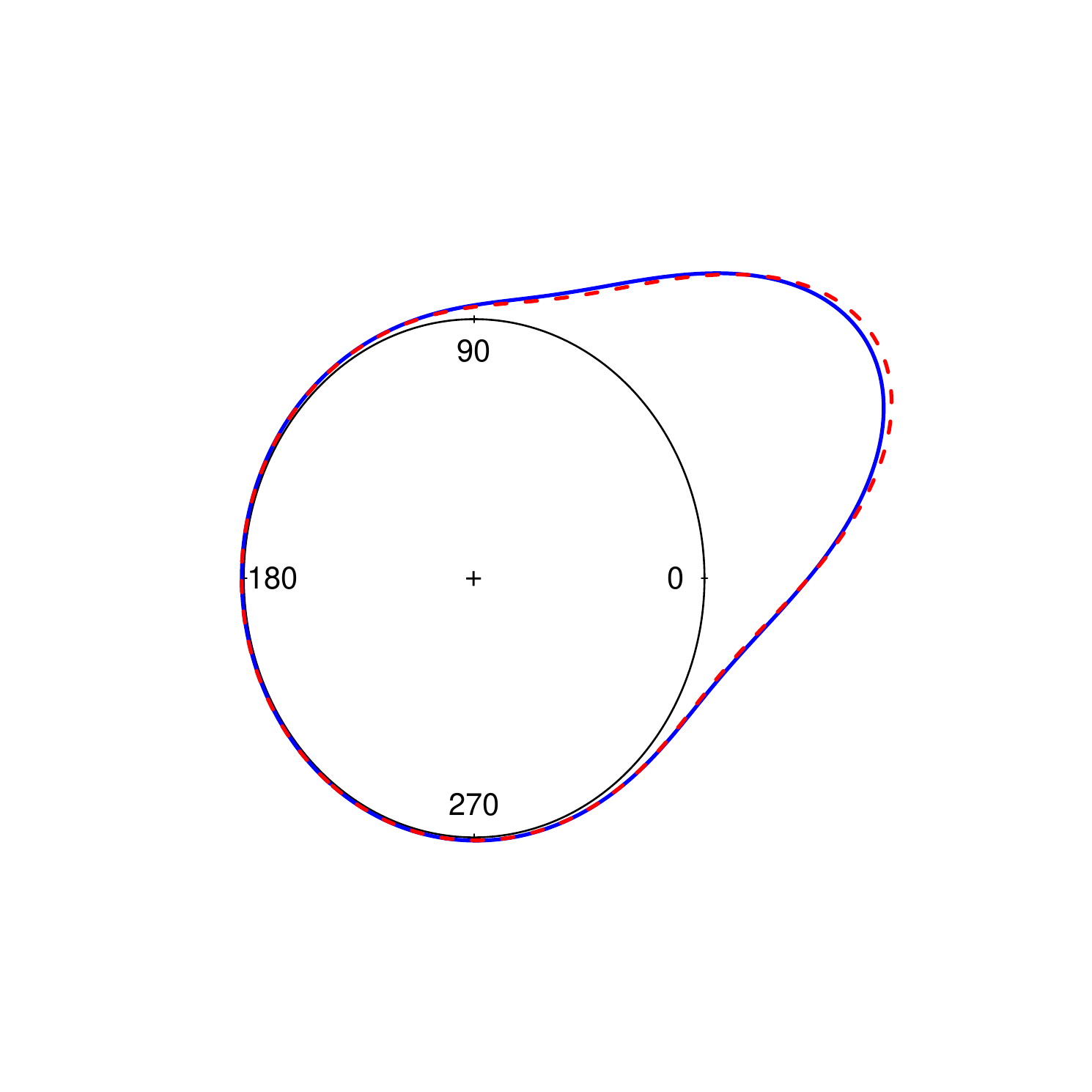} 
		\label{Ankle_den:subfig6}
	}
	\caption{Comparison of posterior predictive densities of the angle of ankle extension of the affected leg of an orthopaedically impaired patient (solid line) and that of the dominating leg of a healthy individual (broken line), keeping the angle of ankle extension of the other leg fixed at the same level.}
	\label{Ankle_den}
\end{figure}

Based on the results, it can be seen from Figure \ref{cont_beta1} that even the $5\%$ confidence ellipse of the posterior distribution of $(Re(\beta_1), Im(\beta_1))$ contains $(0,0)$. This indicates that the rotation model is sufficient to explain the regression for this particular dataset under consideration. To study the difference between healthy and orthopaedically impaired person, we plot the posterior density of $\theta_{\gamma}=\arg(\gamma)$ in Figure \ref{den_gam}. It is clearly visible that the bulk of the distribution is near 0. This is also supported from Figure \ref{gam_test}, and by the fact that $95 \%$ HPD credible interval of $\theta_{\gamma}=\arg(\gamma)$ includes $0$ (see Appendix A of the Supplementary Material). This indicates that the angle of ankle extension of the affected leg of an orthopaedically impaired patient is not significantly different from that of the non-dominating leg of a healthy person when the angle of extension of the other leg are fixed at the same level. Moreover, the $95 \%$ HPD credible interval of $\theta_{01}$ also contains zero, which indicates no significant difference between the angles of ankle extension for both legs for healthy individuals as well as the orthopaedically impaired patients. This is also visible from the posterior predictive distributions of the angle of ankle extension of the affected/non-dominating leg where the modes of the posterior predictive distributions are very close to the respective angle of extension of the healthy/dominating leg (see Figure \ref{Ankle_den}).

%\be
%Y^*=\beta_0 \frac{X^{\as t}+\gamma^{Z_i}\beta_1}{1+\overline{\gamma^{Z_i}\beta_1} X^*}\epsilon_1,
%\label{reg1}
%\ee
%\be
%X^*=b_0 \epsilon_2,
%\label{reg2}
%\ee

%where $Z_i=1$ if the $i$-th observation is coming from an orthopaedically impaired individual and $Z_i=0$ if the $i$-th observation is coming from a healthy person. Here, $\gamma \in \mathbb{C}$. 
%In the table, $\theta_{\gamma}=\arg(\gamma)$.

%\begin{figure}
%\centering
%\begin{tabular}{ccc}
%\text{\scriptsize Treatment vs Control when covariate is $\ang{5}$} & \text{\scriptsize Treatment vs Control when covariate is $\ang{15}$} & \text{\scriptsize Treatment vs Control when covariate is $\ang{25}$} \\
%\includegraphics[width=4cm,height=4cm]{trt_vs_con_52.pdf} &
%\includegraphics[width=4cm,height=4cm]{trt_vs_con_152.pdf} &  	\includegraphics[width=4cm,height=4cm]{trt_vs_con_252.pdf}\\
%\end{tabular}
%\caption{Density plots for posterior predictive distribution  based on three different covariate values for Treatment(blue) and control (red).}
%\label{PPPR}
%	\end{figure}

\section{Discussion}\label{conc}		
In this paper, a Bayesian methodology has been developed for 
a circular-circular regression model with point-accumulation in covariate and response, unlike the existing frequentist methods that only model the cases with point-accumulation in the response variable. Circular-circular regression is not well studied from a Bayesian perspective. This paper makes an attempt in that direction and possibly for the first time Bayesian estimation is proposed for the M\"obius transformation based circular-circular regression model. 
%Our proposed model fits the data well compared to existing models that ignore zero-inflation in the covariates. 
The proposed method is applied to analyse two real datasets on post operative astigmatism and abnormal gait due to orthopaedic impairments. Our analysis provides interesting insights and predictions associated with the recovery of patients. Findings from the analysis of cataract surgery data may help to reduce post-operative trauma and improve the patient's experience after cataract surgery. Similarly, bio-engineers can exploit the analytical findings from abnormal gait data to develop efficient prosthesis.

In general, the methodology is applicable for conventional circular-circular regression with or without point-accumulation in response and/or covariate. Also, one can implement the proposed methodology with other choices of link function and/or choices of angular error distributions. As the latent variables involved in the modeling are continuous, one can easily modify the proposed methodology 
for a discrete response variable with multiple points of accumulation. Therefore, the scope of the proposed method goes far beyond the particular case studies under consideration. For example, one can model wind direction data where the recorded measurements are among some pre-specified discrete directions (e.g. North, West, East, and South). Although we have not  considered missing data in our current
analysis, a simple data augmentation technique has to be incorporated into the proposed methodology when the missingness is ignorable. Another possible direction of future research is to develop a bias-reduced estimation methodology for zero-inflated circular-circular regression using the approach proposed by \citet{schwartz16}. In some real-life scenarios, a circular response may depend on both linear and circular covariates. It will be an interesting problem to model such data and develop associated estimation methodology. As the M\"obius transformation based circular-circular regression model is not readily extendable to the case with linear covariates, it can be considered as future work.

\section*{\small Acknowledgement}
The work of Dr. Prajamitra Bhuyan was supported in part by the Lloyd’s Register Foundation programme on data-centric engineering at the Alan Turing Institute, UK. The authors are thankful to Prof. Anup Dewanji, Dr. Arnab Chakraborty, Prof. Debasis Sengupta, Dr. Jayabrata Biswas, Dr. Sourabh Bhattacharya, and Mr. Sudipta Kundu for many helpful comments and suggestions.
\bibliographystyle{apalike}
\bibliography{Prajamitra}

\end{document}